\documentclass[a4paper,fleqn]{cas-sc}
\graphicspath{{.}}

\usepackage[numbers]{natbib}
\usepackage{graphicx}
\usepackage{float}
\usepackage{subfig}
\usepackage{indentfirst}
\usepackage{listings}
\usepackage{algorithm}
\usepackage{algorithmic}
\usepackage{amssymb}
\usepackage{amsmath}
\usepackage{booktabs}
\usepackage{multirow}
\usepackage{hyperref}
\usepackage{caption}
\def\tsc#1{\csdef{#1}{\textsc{\lowercase{#1}}\xspace}}
\tsc{WGM}
\tsc{QE}

\begin{document}
\let\WriteBookmarks\relax
\def\floatpagepagefraction{1}
\def\textpagefraction{.001}

\shorttitle{SimCLF: A Simple Contrastive Learning Framework for Function-level Binary Embeddings}  

\shortauthors{Sun Ruijin et al} 

\title {SimCLF: A Simple Contrastive Learning Framework for Function-level Binary Embeddings} 
\tnotemark[1] 


%

\author[1]{Sun Ruijin}


\ead{sunruijin6106@163.com}


\credit{<Credit authorship details>}

\affiliation[1]{organization={Army Engineering University of PLA},
      addressline={}, 
      city={Nanjing},
      postcode={210007}, 
      state={},
      country={China}}


\author[2]{Guo Shize}

\ead{nsfgsz@126.com}


\credit{}

\affiliation[2]{organization={National Computer Network and Information Security Management Center},
	addressline={}, 
	city={Beijing},
	postcode={100029}, 
	state={},
	country={China}}
\author[3]{Guo Jinhong}

\ead{guojinhong@sjtu.edu.cn}

\author[1]{Li Wei}
\ead{liwei@aeu.edu.cn}

\author[1]{Zhan Dazhi}

\ead{zhangaga93@aeu.edu.cn}

\credit{}

\affiliation[3]{organization={Shanghai Jiao Tong University},
	addressline={}, 
	city={Shanghai},
	postcode={200240}, 
	state={},
	country={China}}

\author[4]{Guo Xi}
\ead{xiguo@ustb.edu.cn}


\credit{}
\affiliation[4]{organization={University Of Science \& Technology Beijing},
	addressline={}, 
	city={Beijing},
	postcode={100083}, 
	state={},
	country={China}}

\author[1]{Sun Meng}


\ead{sunmengccjs@163.com}

\author[1]{Pan Zhisong}

\cormark[1]

\ead{hotpzs@hotmail.com}


\credit{<Credit authorship details>}





\begin{abstract}
			Function-level binary code similarity detection is a crucial aspect of cybersecurity. It enables the detection of bugs and patent infringements in released software and plays a pivotal role in preventing supply chain attacks. A practical embedding learning framework relies on the robustness of the assembly code representation and the accuracy of function-pair annotation, which is traditionally accomplished using supervised learning-based frameworks. However, annotating different function pairs with accurate labels poses considerable challenges. These supervised learning methods can be easily overtrained and suffer from representation robustness problems. To address these challenges, we propose SimCLF: A Simple Contrastive Learning Framework for Function-level Binary Embeddings. We take an unsupervised learning approach and formulate binary code similarity detection as instance discrimination. SimCLF directly operates on disassembled binary functions and could be implemented with any encoder. It does not require manually annotated information but only augmented data. Augmented data is generated using compiler optimization options and code obfuscation techniques. The experimental results demonstrate that SimCLF surpasses the state-of-the-art in accuracy and has a significant advantage in few-shot settings.
\end{abstract}


\begin{highlights}
\item To investigate the robustness of the embeddings, our finding reveals that SOTA models encounter expression degradation issues, and the reasons for these problems are analyzed. 
\item This is the first article that employs unsupervised contrastive learning in Binary Code Similarity Detection. Our model effectively circumvents the labelling difficulties in existing approaches.
\item Our findings indicate that SimCLF outperforms existing models. Additionally, SimCLF is tested on a real-world vulnerability dataset and it has achieved flawless results.
\item SimCLF demonstrates greater efficiency in few-shot settings.
\item We hope that our models will inspire others. We have released the code of SimCLF publicly available at \url{https://github.com/iamawhalez/fun2vec}. 
\end{highlights}


\maketitle 
{SimCLF: A Simple Contrastive Learning Framework for Function-level Binary Embeddings}
\section{Introduction
\label{sec_instr} }
Function-level binary code similarity detection (BCSD) is used to analyze whether two given binary code snippets are similar in semantics. It plays an important role in various tasks of cyberspace security \citep{haq_survey_2019}, such as bug search \cite{10.1145/3238147.3238199}, plagiarism detection \cite{7823022}, vulnerability discovery \cite{10.1007/978-3-540-88625-9_16}, malware detection \cite{10.1145/1653662.1653736}, and malware clustering \cite{10.5555/2535461.2535485}. Several researchers \citep{pei_TREX_nodate}\citep{li_palmtree_2021}\citep{massarelli_safe_2019}\citep{noauthor_semantic_nodate}\citep{xu_neural_2017} have demonstrated promising results in the analysis of binary code similarity. There are many challenges associated with BCSD. The foremost one is the huge number of the functions to be detected. As the Internet era progresses, new functions and new applications emerge in an endless stream, and the number of binary files increases rapidly \citep{massarelli_safe_2019}. In addition to the huge number, the architectures of the devices are becoming more and more diverse. Besides the traditional x86 with Windows, there are a variety of other operating systems (Ubuntu, IOS, Android etc.) and CPU types (Arm, MIPS etc.). When new software are developed, people often do not start from scratch. They may reuse existing code snippets or open source libraries. If there is a bug in the original library file, that bug will be present in all programs that reuse it. It is impossible to manually identify all the library files that software contains in the face of massive data. Therefore, there is an urgent need for automated methods based on the semantics of binary files. 

Despite the highly practical and academic value of BCSD, progress has been very slow because of the lack of quality labelled datasets in the BCSD field. Annotating binary function pairs is a challenging task, and security analysts often face difficulty in obtaining the source code of the software being analyzed for various reasons. Additionally, compiled binary files do not contain high-level language components such as variable names and data structures, as most of the program semantics are lost during the compilation process. Binaries generated from the same high-level language can undergo significant changes when compiled using different compilers, optimization options, and on different CPU architectures. Additionally, hackers often use obfuscation techniques to distort binary code, making BCSD an increasingly difficult task. Unlike labelling in Natural Language Processing (NLP), labelling binary code requires many specialists and a lot of hard work. All of these issues contribute to difficulties in annotating binary code. Despite overcoming the aforementioned challenges, labelling different function pairs accurately remains challenging. Supervised learning-based methods typically use distance functions, such as edit distance and cosine similarity \citep{massarelli_safe_2019}, to label function pairs. The distance between the two similar functions could be labelled as $\left \{ 1\right \}$ and the two different functions as $\left \{ -1 \right \}$ or $\left \{ 0 \right \}$. But the mutual information between any pair of the function could not be either $\left \{ -1 \right \}$ or $\left \{ 0 \right \}$. For example, we may label $\mathit{Recv()}$ and $\mathit{RecvAndWrite()}$ as $\left \{ -1 \right \}$, because they are not the same function. However, in reality, these two functions share significant similarities, the ground truth should be a score between $-1$ and $1$. Finding an appropriate number for labelling different function pairs can be challenging. If we use $\left \{ -1 \right \}$ to label them, it can lead to overtraining which causes a slow convergence.

\begin{figure}
	\begin{minipage}[t]{0.45\linewidth}
		\centering
		\includegraphics[scale=0.37]{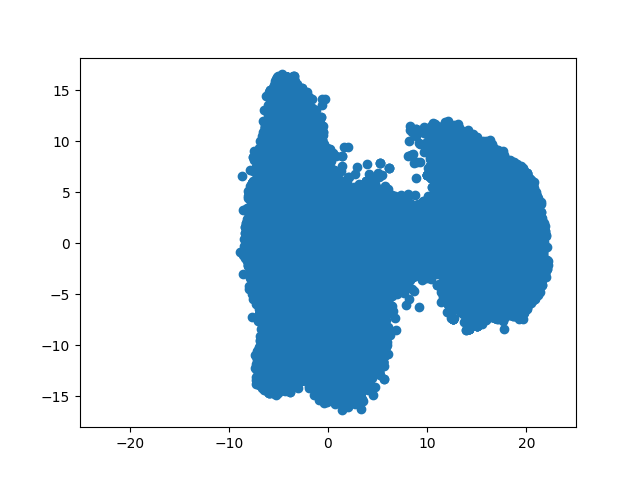}
		\centerline{(a) A pretrained Transformer model.}
	\end{minipage}%
	\begin{minipage}[t]{0.45\linewidth}
		\centering
		\includegraphics[scale=0.37]{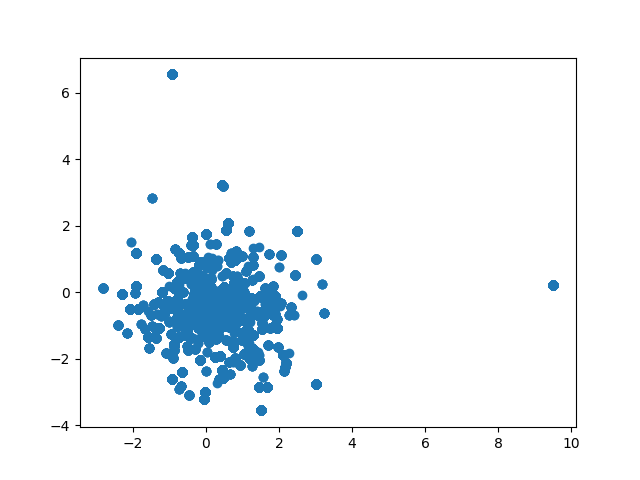}
		\centerline{(b) Word2Vec on assembly codes.}
		\label{first}
	\end{minipage}%
	\caption{The figure displays two-dimensional visualizations of embeddings. Figure (a) shows the visualization of embeddings generated from a pre-trained transformer model called TREX \citep{pei_TREX_nodate}. On the other hand, Figure (b) shows the visualization of embeddings trained using Word2Vec \cite{10.5555/2999792.2999959}. }
	\label{fig:corner}
\end{figure}
 A reliable framework for binary code similarity analysis depends on a robust vector representation of assembly code. While current methods have achieved state-of-the-art (SOTA) results in BCSD \citep{pei_TREX_nodate}\citep{li_palmtree_2021}\citep{yu_order_2020}\citep{noauthor_semantic_nodate}, many pretrained models face the issue of representation degeneration. To address this, Singular Value Decomposition (SVD) is utilized to approximate the learned embeddings of the SOTA model TREX \citep{pei_TREX_nodate}. As shown in Figure \ref{fig:corner}, the rank-2 approximation of the embeddings for all datasets that used in the finetuning process is plotted. The detail of the dataset is described in Section \ref{sec:setups}. These embeddings are compared with those trained by Word2Vec \citep{10.5555/2999792.2999959}. The 2D visualization of the representations from the pre-trained model collapses to a corner, while the word embeddings learned from Word2Vec \citep{10.5555/2999792.2999959} are uniformly distributed. 
 
Motivated by the latest developments in the area of contrastive learning \citep{wu_unsupervised_2018}\citep{gao_simcse_2022}\citep{chen_simple_2020}\citep{yan_consert_2021}, we propose a novel binary code semantic difference (BCSD) model at the function level, namely SimCLF. SimCLF is a self-supervised framework that leverages contrastive learning and solely relies on positive examples and sample augmentation techniques. This approach helps to avoid the over-training problem that may arise from the inaccurate labelling of negative pairs. Additionally, the experimental findings demonstrate that SimCLF is capable of eliminating the anisotropy \citep{ethayarajh_how_2019} caused by the pre-trained model. In terms of accuracy, it surpasses existing models such as TREX \cite{pei_TREX_nodate} and jTrans \cite{wang2022jtrans}. In summary, our contributions are as follows:
\begin{itemize}
	\item To investigate the robustness of the embeddings generated by SOTA pre-training models, several sets of metrics are developed. The findings suggest that these models suffer from expression degradation issues, and the underlying causes of these problems have been analyzed.
	\item We propose SimCLF, a simple contrastive learning framework of function-level binary embeddings. As far as we know, this is the first article that utilizes the unsupervised contrastive learning model in BCSD. SimCLF avoids the labelling difficulties that arise in supervised models.
	\item The model is subjected to rigorous evaluation across multiple datasets, and its performance is compared against the existing SOTA approaches. The results reveal that SimCLF surpasses the performance of its counterparts. Furthermore, SimCLF is applied to a real-world vulnerability dataset, where it has achieved perfect results.
	\item Our model is more efficient in few-shot settings, as it achieves better performance than TREX \citep{pei_TREX_nodate} with only 2,048 binary pairs. In comparison, TREX requires about 89,809 labelled binary pairs.
	\item It is hoped that SimCLF will inspire others in the field. To facilitate further research, the code for SimCLF have been released at \url{https://github.com/iamawhalez/fun2vec}. 
\end{itemize}

\section{Background and Related Work}
In this section, we provide a brief introduction to the definition of BCSD and related work in the field.
\subsection{Definition of BCSD}
The vast majority of software development is characterized by the practice of code reuse, it is rare for developers to initiate a project from scratch. As a result, a significant number of code clones are likely to exist within the underlying assembly code. An efficient BCSD engine can greatly reduce the burden of the manual analysis process involved in reverse engineering. By exploiting the large amount of available binary data, BCSD can cater to the information requirements of downstream tasks. The definition of BCSD can be described as follows : given two pieces of binary code, without any supplementary information such as source code or compilation details, the task is to determine whether the two pieces of code are similar. In this context, similarity refers to semantic consistency, but not necessarily the same characters. In other words, if similar binary functions belong to different architectures or are compiled using different optimizations, they will still behave the same way when they are provided with the same input.
BCSD methods can be broadly categorized into two distinct groups: feature-based methods and embedding-based methods. 
\subsection{BCSD Based on Features}
Traditional BCSD methods heavily rely on specific features or structures. The earliest tool for comparing binary file similarity, named EXEDIFF \citep{Baker99compressingdifferences}, was released in 1999 and was used to determine the patch information of different versions of executable files. These methods primarily relied on disassembled code structure information, and string-based detection techniques are the most fundamental methods \citep{514697}\citep{10.1145/956750.956759}. String-based techniques use string or string edit distance to detect duplicates. Another commonly used feature is $n$-$grams$ wherein sliding windows of size $n$ in bytes are considered as content \citep{khoo_rendezvous_2013}. Many articles adopted a graph-based approach for BCSD. Graph-based methods can be broadly categorized into those that utilize Program Dependency Graphs (PDGs) and those that use Control Flow Graphs (CFGs). A PDG is a graphical representation of the data and control dependencies within a procedure \citep{Liu06gplag:detection}. It was used in program slicing at first \citep{6956589}. More people use CFG (Control Flow Graph), such as \citep{7163056} and \citep{10.1145/2664243.2664269}. These methods transform binary code into CFGs and use graph analysis algorithms to determine similarity.
\subsection{BCSD Based on Embeddings}
Embedding is the transformation of discrete variables into a high-dimensional continuous space. In embedding-based solutions, functions are not compared based on the function itself, but instead on the learned embedding of functions. First, binary functions are transformed into multi-dimensional vector representations (embeddings). These embeddings are then compared using simple geometric operations \citep{massarelli_safe_2019}. Functions with similar semantics have embeddings that are closer to each other in the high-dimensional space.

In literature \citep{ding_asm2vec_2019}, the authors proposed a function embedding solution called Asm2Vec. Asm2Vec transforms the assembly code into embeddings by Word2Vec\citep{10.5555/2999792.2999959}. The function representations are learned by using code embeddings based on the PV-DM model \citep{le_distributed_nodate}. Some articles employ graph embedding methods. For instance, \cite{xu_neural_2017} use a neural network-based approach to learn the embedding of an Attributed Control Flow Graph (ACFG). Initially, SAFE \citep{massarelli_safe_2019} employs Word2Vec to convert the instructions into embeddings. A GRU Recurrent Neural Network (GRU RNN) is then used to capture the sequential interaction of the instructions. 

The success of pure attention networks in the NLP domain such as Transformer \citep{10.5555/3295222.3295349} and BERT \citep{devlin-etal-2019-bert} has inspired the adoption of Transformer-like structures in BCSD. All of these articles \citep{noauthor_semantic_nodate}\citep{li_palmtree_2021}\citep{pei_TREX_nodate}\citep{yu_order_2020} employ architectural structures that are rooted in the pure attention pre-training model, yielding remarkable outcomes. In \citep{yu_order_2020}, the authors adopt a method based on BERT, where they pre-train the binary code using BERT at both the one-token and one-block levels. Then Convolutional Neural Networks (CNN) are used on adjacency matrices to extract the order information. PalmTree \citep{li_palmtree_2021} utilizes three pre-training tasks to capture various characteristics of assembly language and generate general-purpose instruction embeddings. These pre-training tasks are performed through self-supervised training on large-scale unlabelled binary corpora. TREX \citep{pei_TREX_nodate} develops a novel neural architecture, a hierarchical Transformer, which can learn execution semantics from micro-traces during the pre-training phase. jTrans\cite{wang2022jtrans} is the first solution that embeds control flow information of binary code into Transformer-based language models. However, none of these papers have investigated the quality of the embedding that they've learned. Our experiments indicated that the issue of representation collapse exists in code embedding based on pre-trained models.
\section{The Basic Structure of the Model}
In this section, the basic structure of SimCLF model is presented, and the technical approach behind it is discussed. We start by describing the general algorithm, then briefly introduce the major parts of the model. Throughout this paper, the following mathematical notations are adopted. The element or binary function in the original dataset $X$ is denoted as $x$, while $f(\cdot)$ represents a deep neural network-based encoder. We primarily use BERT-like pretrained model TREX \citep{pei_TREX_nodate} and jTrans\cite{wang2022jtrans} as the encoder. We aim at fine-tuning the encoder $f(\cdot)$ to make the function representation more robust. $Trans(\cdot)$ is the data augmentation function. $z$ is the learned embedding of the function $x$. $\tau $ is a temperature parameter. $d(\cdot)$ is a distance function, $\mathcal{L}(\cdot)$ is a loss function.
\subsection{General Framework}
\begin{figure}[htbp]
	\centering 
	\includegraphics[scale=0.24]{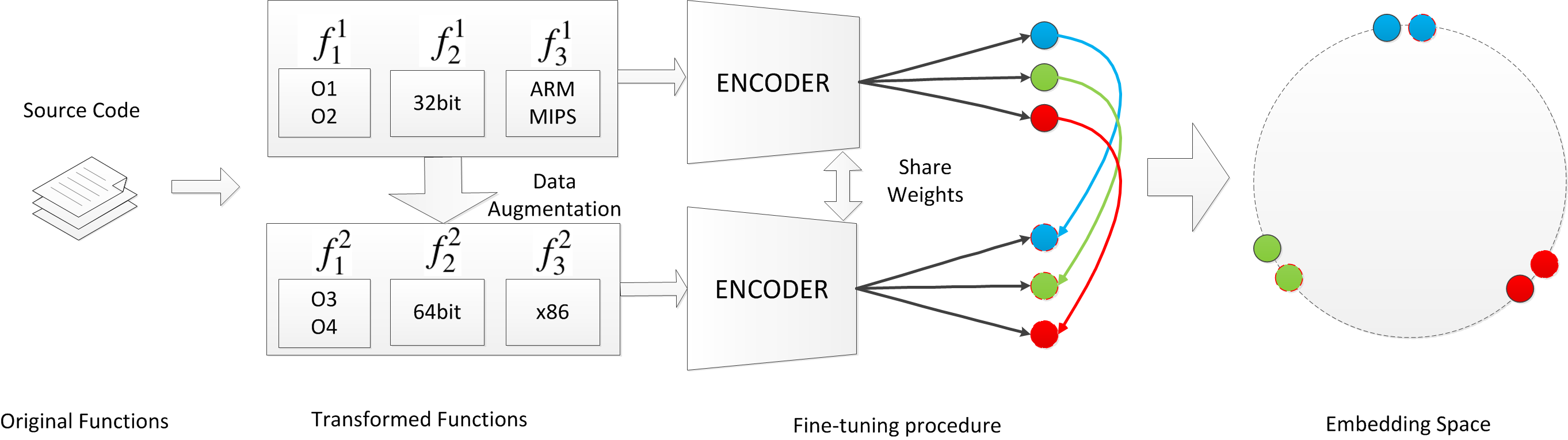} 
	\caption{ The proposed unsupervised learning method with siamese networks follows a framework where input function pairs are projected into low-dimensional, normalized embeddings through the use of a DNN-based encoder. After undergoing the fine-tuning process, the feature representations of the same function instance with different data augmentations are invariant, while the embeddings of different function instances are distributed.} 
	\label{fig1}
\end{figure}

Inspired by the advancements in contrastive learning and natural language processing, we propose SimCLF, which aims to address the two challenges posed in the Section \ref{sec_instr}. The model consists of a siamese network, which implements the proposed encoder, and a contrastive learning model as shown in Figure \ref{fig1}. By employing the unsupervised approach, the manual labor involved in labelling different function pairs can be circumvented, while simultaneously enhancing the quality of the embedding to a significant extent. SimCLF consists of three parts.
\begin{itemize}
	\item A data set $X$ and the sample augmentation module $Trans(\cdot)$. 
	\item A DNN-based encoder $f(\cdot)$ that projects input function pairs into low-dimensional. 
	\item A contrastive learning module that is used to optimize the encoder, which consists of a distance function $d(\cdot)$ and a loss function $\mathcal{L}(\cdot)$.
\end{itemize}

Given a function $x_{i}$ in the original dataset $X$, the augmented data are produced by passing the original $x_{i}$ to the data augmentation module $Trans(\cdot)$. Two augmented binary functions, $x_{i}^{1}$ and $x_{i}^{2}$, are generated. Then, the two embeddings $z_{i}^{1}$ and $z_{i}^{2}$ of the function-pair are produced. 
$z_{i}^{1}=f(x_{i}^{1})$ and $z_{i}^{2}=f(x_{i}^{2})$ where $f$ is the representation learning module, are also known as the encoder. 
After generating augmented data through the data augmentation module, the contrastive learning module is used to optimize the encoder. A key concept introduced here is non-parametric instance discrimination \citep{wu_unsupervised_2018}. We treat each function instance as a distinct class of its own and train a classifier to discriminate between individual instance classes. In other words, given any function pair $x_{i},x_{j}$, if $x_{i}$ and $x_{j}$ are semantically different, it is believed that they are members of different classes. The distance function $d(i,j)$ between the embeddings of $x_{i}$ and $x_{j}$ should be greater than a threshold value. If $x_{i}$ and $x_{j}$ are semantically the same, it is believed that they belong to the same classes. And the distance function $d(i,j)$ between the embeddings of $x_{i}$ and $x_{j}$ should be less than a threshold value.
\subsection{Data Augmentation Strategies}
The data augmentation module plays a crucial role in contrastive learning. In the realm of binary code, binary codes with the same meaning may have completely different assemblys on different devices. The data augmentation module is capable of generating a diversity of assemblys. Given a binary function code, different binaries are produced with the same semantics, which may be suitable for different architectures or compiled with different optimizations or obfuscation techniques. In this paper, three different data augmentation strategies are explored, namely cross-architecture, cross-optimization, and obfuscation, to generate augmented instances for the contrastive learning process.\\
\textbf{Cross-architecture}. In general, the CPU platforms are randomly selected on which the code will run. The two main parameters in this process are the CPU architecture and width. The architectures include x86-32, x86-64, arm-32, arm-64, mips-32, and mips-64, covering the most common mainstream CPU types.\\
\textbf{Cross-optimization}. The assembly codes produced by different compilation options for the same function written in the same high-level language are also very different. Given a specific function, it will be compiled into different architectures with 4 optimization options,$i.e.$ ,$O0,O1,O2$ and $O3$ and 2 compilers $GCC$-$5.4$,$GCC$-$7.5$.\\
\textbf{Obfuscation}. Hackers often use code obfuscation techniques to prevent security personnel and tools from analyzing their code. To simulate this scenario, Hikari \citep{obf} is used to generate obfuscated augmented instances for the contrastive learning process. Several common obfuscation techniques are applied, including Pseudo Control Flow, Instruction Substitution, Basic Block Splitting, and others. 
\subsection{Function Representation Learning Module}%
In this section, the binary function representation learning module that we adopted are briefly introduced. Given a specific pair of binary functions, they are converted into continuous vectors in high dimensional space by the function representation learning module. The similarity of the binary pair is then calculated according to the distance of two vectors. SimCLF allows for any network architecture for encoding. Two model TREX\citep{pei_TREX_nodate} and jTrans \citep{wang2022jtrans} are chosen based on the following considerations. Firstly, these two model are the currently SOTA models. Pre-training based models are currently the most promising models, and many articles \citep{pei_TREX_nodate}\citep{li_palmtree_2021}\citep{yu_order_2020}\citep{noauthor_semantic_nodate} have adopted it. Secondly, TREX and jTrans have open-sourced their codes, unlike some other models which are not fully disclosed. This allowed us to easily incorporate it into our framework and adapt it to suit our needs. 

TREX is a pre-trained language model. It can generate embeddings for each token in an assembly function. TREX could encode the input functions and obtain their respective embeddings in the vector space. The function representations are obtained by performing average pooling of the token embeddings at the last layer.

jTrans is based on the BERT architecture. It capture control-flow information, by sharing parameters between token embeddings and position embeddings for each jump target of instructions. jTrans is trained by unsupervised learning tasks to learn the semantics of instructions and control-flow information of binary functions. Finally, the BertPooler layer is used to obtain the function embedding.
\subsection{Contrastive Learning Module}
In this section, a detailed description of the contrastive learning module is provided, which consists of two components: a distance function and a loss function. To evaluate the similarity of function pairs, the cosine distance function $sim()$ are used. 
Given a pair of functions $x_{1} $ and $x_{2}$, $\boldsymbol{z}_{1}=\vec{f(x_{1})}$ and $
\boldsymbol{z}_{2}=\vec{f(x_{2})}$ are the respective embedding vectors produced by the encoder. $w$ is the dimension of embeddings. Cosine similarity are used as the distance metric $d(\cdot)$, which is calculated using the following formula:
\begin{equation}
d(\boldsymbol{z}_{1},\boldsymbol{z}_{2})=sim(f(x_{1}),f(x_{2}))=\frac{\vec{f(x_{1})} ^{\top} \vec{f(x_{2})}}{\left \| \vec{f(x_{1})}\right \| \left \| \vec{f(x_{2})}\right \| } =\frac{\sum_{i=1}^{w} \left (\vec{f(x_{1}) }\left [ i \right ] \cdot \vec{f(x_{2}) }\left [ i \right ]  \right ) }{\sqrt{\sum_{i=1}^{w} \vec{f(x_{1})}\left [ i \right ]^{2} } \cdot \sqrt{\sum_{i=1}^{w} \vec{f(x_{2})}\left [ i \right ]^{2} } } ,
\end{equation}
and $\vec{f(x)}\left [ i \right ]$ indicates the $i$-$th$ component of the embedding vector of the function $x$. The cosine distance between two function embeddings ranges from $[-1,1]$, where a distance of 1 indicates that the functions are identical and a distance of -1 indicates that they are completely different. 

If there are $n$ different functions, $f(x_{i}),i\in [1,n]$ is the learned representation of the $i$-$th$ function. We enforce $\left \|\boldsymbol{z} \right \|= 1$ via an $L2$-$normalization$ layer. Following the approach described in reference \citep{wu_unsupervised_2018}, the probability are formulated that any embedding $z$ belongs to the $i$-$th$ function as an instance discrimination objective using the softmax-like criterion,

\begin{equation}
P(\boldsymbol{z}_{i} \mid \boldsymbol{z})=\frac{\exp \left( \mathbf{z}_{i}^{T} \mathbf{z} \right)}{\sum_{j=1}^{n} \exp \left(\mathbf{z}_{j}^{T} \mathbf{z}\right)} .
\end{equation} 
The distance function is substituted into equation 2, where $sim(z_{i}, z)$ denotes the dot product between L2 normalized $z_{i}$ and $z$. 
\begin{equation}
P(z_{i} \mid \mathbf{z})=\frac{\exp \left( sim(z_{i},z) \right)}{\sum_{j=1}^{n} \exp \left(sim(z_{i},z)\right)} ,
\end{equation}
In the context of the given scenario, $P(z_{i} \mid \mathbf{z})$ represents the probability that the embedding $z$ belongs to class $i$. A temperature parameter $\tau $ is incorporated to regulate the concentration level \citep{hinton2015distilling}. The probability $P(z_{i} \mid \mathbf{z})$ is modified as follows:
\begin{equation}
P(z_{i} \mid \mathbf{z})=\frac{\exp \left(sim(z_{i},z)/ \tau)\right)} {\sum_{j=1}^{n} \exp \left(sim(z_{i},z)/\tau\right)} ,
\label{eq_tau}
\end{equation}

Our objective is to maximize the joint probability $\prod_{i=1}^{n} P(z_{i} \mid \mathbf{z})$, which is equivalent to minimizing the negative log-likelihood $-\sum_{i=1}^{n}\log{P(z_{i} \mid \mathbf{z})} $.
The loss function becomes the NT-Xent (the normalized temperature-scaled cross-entropy loss) \citep{chen_simple_2020}, which needs to be minimized during the fine-tuning process.

During each training epoch, a particular function $x_{i}$ in the batch of $N$ is selected. And two augmented functions $x_{i}^1$ and $x_{i}^2$ of the same semantic are randomly generated. This results in a total of $2N$ data points denoted as $z_{n},n\in [1,2N]$. In our approach, negative examples are not explicitly sampled. Instead, when given a positive pair $x_{i}^1$ and $x_{i}^2$, the corresponding vectors are $z_{2i}=f(x_{i}^1)$ and $z_{2i+1}=f(x_{i}^2)$. The remaining $2(N-1)$ augmented examples within the minibatch are treated as negative examples. The vector $z_{2i}$ is similar to $z_{2i+1}$, and distinct from all the other $2(N-1)$ augmented examples. Consequently,s the loss function of $x_{i}^{1}$ and $x_{i}^{2}$ could be formulated as
\begin{equation}
\mathcal{L} _{x_{i}^1}=-log \frac{exp(sim(z_{2i}, z_{2i+1})/\tau )}{{\textstyle \sum_{k=1}^{2N }} \mathbb{L}_{[ k\ne 2i]} exp(sim(z_{2i}, z_{k})/\tau)},
\mathcal{L} _{x_{i}^2}=-log \frac{exp(sim(z_{2i+1}, z_{2i})/\tau )}{{\textstyle \sum_{k=1}^{2N }} \mathbb{L}_{[ k\ne 2i+1]} exp(sim(z_{2i+1}, z_{k})/\tau)} .
\end{equation} 
Then the loss function for a positive pair of examples $x_{i}^{1}$ and $x_{i}^{2}$ is defined as:
\begin{equation}\mathcal{L} _{x_{i}} =1/2(\mathcal{L} _{x_{i}^1}+\mathcal{L} _{x_{i}^2}) =-1/2 \cdot \left ( log \frac{exp(sim(z_{2i}, z_{2i+1})/\tau )}{{\textstyle \sum_{k=1}^{2N }} \mathbb{L}_{[ k\ne 2i]} exp(sim(z_{i}, z_{k})/\tau)}+log \frac{exp(sim(z_{2i+1}, z_{2i})/\tau )}{{\textstyle \sum_{k=1}^{N }} \mathbb{L}_{[ k\ne 2i+1]} exp(sim(z_{i}, z_{k})/\tau)} \right ) .
\end{equation} 
The final loss $\mathcal{L}$ is computed over all positive pairs, where $\mathcal{L} =\frac{1}{N} \sum_{k=1}^{N} \mathcal{L} _{x_{i}}$. The employed loss function $\mathcal{L}$ effectively maximizes the dissimilarity between a function and its augmented data, while simultaneously minimizing the cosine distance between distinct functions. Through the minimization of this loss function, we are able to achieve instance classification without relying on labeled samples of different types. The complete algorithm outlining the entire process is presented in Algorithm \ref{alg:Framwork}.
\begin{algorithm}[htb] 
	\caption{ Framework of the Contrastive Learning Module SimCLF.} 
	\label{alg:Framwork} 
	\begin{algorithmic}[1] 
		\REQUIRE ~~\\ 
		The set of positive samples for the current batch, $\left \{ x_{i}\right \} _{1}^{n}$;
		the temperature constant $\tau$;
		a DNNs structure of the encoder $f(\cdot)$; the data augmentation function $Trans()$;
		\ENSURE ~~\\ 
		
		\FOR {each $x_{i} \in$ sampled minibatch $\left \{x_{i}\right \}_{1}^{n}$ }
		\STATE Initialize $x_{i}^1$,$x_{i}^2$ with the augmentation function $x_{i}^1,x_{i}^2 =Trans(x_{i})$
		\STATE Encode them with the encoder $z_{2i}=f(x_{i}^1)$, $z_{2i+1}=f(x_{i}^2)$

		\STATE For each pair of functions embeddings, calculate their cosine distance, $sim(z_{2i}, z_{2i+1}) =\frac{z_{2i}^{\top } z_{2i+1} } {\left \|z_{2i} \right \| \left \| z_{2i+1} \right \| } $
		\STATE The loss function for a pair of functions is defined as \begin{equation}
  \mathcal{L} _{x_{i}} =1/2(\mathcal{L} _{x_{i}^1}+\mathcal{L} _{x_{i}^2}) =-1/2 \cdot \left ( log \frac{exp(sim(z_{2i}, z_{2i+1})/\tau )}{{\textstyle \sum_{k=1}^{2N }} \mathbb{L}_{[ k\ne 2i]} exp(sim(z_{i}, z_{k})/\tau)}+log \frac{exp(sim(z_{2i+1}, z_{2i})/\tau )}{{\textstyle \sum_{k=1}^{N }} \mathbb{L}_{[ k\ne 2i+1]} exp(sim(z_{i}, z_{k})/\tau)} \right )\nonumber ,
  \end{equation}
		\STATE The final loss is calculated over all positive pairs,$\mathcal{L} = \frac{1}{N} \sum_{k=1}^{N} \mathcal{L} _{x_{i}}$ 
		\STATE Minimize $\mathcal{L}$ to update networks $f$
		\ENDFOR
		\RETURN finetuned network $f$; 
	\end{algorithmic}
\end{algorithm}

\subsection{Why does SimCLF work}
SimCLF is the first paper to apply unsupervised contrastive learning in the field of BCSD. Unsupervised contrastive learning has garnered remarkable success, whereas the underlying mechanism of contrastive loss has received comparatively less scholarly attention. 
Taking inspiration from
\cite{10.5555/3524938.3525859}\cite{gao_simcse_2022}\cite{9577669}, our objective is to offer a comprehensive investigation into the underlying principles of SimCLF.
First, SimCLF demonstrates sensitivity to hard examples. Automatic selection of these hard examples can make training more effective and efficient \cite{7780458}. Through an analysis of the gradients pertaining to the SimCLF loss function, it has been discovered that SimCLF is sensitive to hard examples. More precisely, the gradients associated with the two negative pairs $m$ and 
$n$, where $m,n\ne 2i,2i+1$ are formulated as:
\begin{equation}
\frac{\partial({L(x_{i}^1)}) }{\partial(sim(z_{2i},z_{m}))}=\frac{1}{\tau}  \frac{exp(sim(z_{2i}, z_{m})/\tau )}{{\textstyle \sum_{k=1}^{2N }} \mathbb{L}_{[ k\ne 2i]} exp(sim(z_{2i}, z_{k})/\tau)} ,
\end{equation} 

\begin{equation}
\frac{\partial{L(x_{i}^1)} }{\partial(sim(z_{2i},z_{n}))}=\frac{1}{\tau}  \frac{exp(sim(z_{2i}, z_{n})/\tau )}{{\textstyle \sum_{k=1}^{2N }} \mathbb{L}_{[ k\ne 2i]} exp(sim(z_{2i}, z_{k})/\tau)} ,
\end{equation} 
The observation that the gradients with respect to negative samples are proportional to the exponential term, it suggests that SimCLF exhibits sensitivity to hard examples.
\begin{equation}
\frac{\partial{L(x_{i}^1)} }{\partial(sim(z_{2i},z_{m}))}/\frac{{L(x_{i}^1)} }{\partial(sim(z_{2i},z_{n}))}=\frac{exp(sim(z_{2i},z_{m})/\tau)}{exp(sim(z_{2i},z_{n})/\tau)} ,
\end{equation}

 Second, SimCLF significantly diminishes both alignment and uniformity. In a recent study, \citet{10.5555/3524938.3525859} identified two essential properties pertaining to contrastive learning, namely alignment and uniformity. These properties serve as metrics for evaluating the quality of learned representations. Let $p_{pos}(\cdot)$ denote the distribution of positive pairs. The concept of alignment  aims to quantify the distance between embeddings of positive paired instances.

\begin{equation}
\mathcal{A}lignment =\underset{x_{i}^1,x_{i}^2\in p_{pos}(\cdot)}{\mathbb{E}}{}\left \| f(x_{i}^1)-f(x_{i}^2) \right \| ,
\label{eq_Alignment}
\end{equation}
Let $p_{data}(\cdot)$ be the negtive data distribution. Uniformity measures how well the embeddings of negative pairs are uniformly distributed. 
\begin{equation}
\mathcal{U}_{niform} =\underset{x, y \sim p_{\text {data }}}{\mathbb{E}} e^{-2{\left \| f(x)-f(y) \right \|}^2 } ,
\end{equation}
Let the loss function of ${x_{i}^1}$ could be formulated as 
\begin{equation}
\mathcal{L} _{x_{i}^1}=-log \frac{exp(sim(z_{2i}, z_{2i+1})/\tau )}{{\textstyle \sum_{k=1}^{2N }} \mathbb{L}_{[ k\ne 2i]} exp(sim(z_{2i}, z_{k})/\tau)} ,
\end{equation}
It can be divided into two parts
\begin{equation}
 \mathcal{L} _{x_{i}^1}=-\frac{1}{\tau}\sum sim(z_{2i}, z_{2i+1}) +  log
{\textstyle \sum_{k=1}^{2N }}\mathbb{L}_{[ k\ne 2i]} exp(sim(z_{2i}, z_{k})/\tau) ,
\end{equation}
$\mathcal{L}=\frac{1}{N}\sum_{i=1}^{N } (\mathcal{L} _{x_{i}^1}+\mathcal{L} _{x_{i}^2})$,Because $\mathcal{L} _{x_{i}^1}$ and $\mathcal{L} _{x_{i}^2}$ is completely symmetrical, we take $\mathcal{L} _{x_{i}^1}$ as an example.

\begin{equation}
\frac{1}{N}\sum_{i=1}^{N } \mathcal{L} _{x_{i}^1}= \frac{1}{N}(-\frac{1}{\tau}\sum_{i=1}^{N } sim(z_{2i}, z_{2i+1}) +  \sum_{i=1}^{N }log
{\textstyle \sum_{k=1}^{2N }}\mathbb{L}_{[ k\ne 2i]} exp(sim(z_{2i}, z_{k})/\tau)) ,
\label{eqlossterm12}
\end{equation}
The first term of the loss function maintains the similarity of positive instances, while the second term acts to separate negative pairs. The loss function is mathematically expressed by the Equation (\ref{eqlossterm12}). As indicated by \cite{10.5555/3524938.3525859},  in the scenario where the number of instances tends towards infinity and $f (\cdot)$ is normalized, the first term is exactly the alignment term as shown in Equation (\ref{eq_Alignment}). Regardless of the constant factors, the second term signifies favorable uniformity \cite{gao_simcse_2022}\cite{DBLP:journals/corr/abs-2111-00743}. Thus, the optimization of the InfoNCE loss facilitates the attainment of both favorable alignment and favorable uniformity.

Third, the data augmentation module plays a crucial role in contrastive learning. Richer data augmentation indicates better downstream performance \cite{DBLP:journals/corr/abs-2111-00743}. SimCLF encompasses an extensive data augmentation methods, including the most common mainstream CPU types, 4 optimization options and different compilers. 
The effectiveness of the model has been theoretically examined, with the rationale behind its efficacy discussed. To further substantiate its effectiveness, the key metrics will be validated in Section \ref{sec:expandeval}.
\section{Experiment and Evaluation
\label{sec:expandeval}}
 The dataset and experimental configuration are introduced in this section. Our evaluation aims to answer the following question:\\
RQ1. Basic statistical laws of assembly code. What are the basic statistical laws of assembly code, and is there an expression degradation problem in the results of the original pre-trained model? Additionally, we investigate whether fine-tuning the model alleviates the expression degradation issue.\\
RQ2. Accuracy of the model. What is the accuracy of our contrastive learning model, and how does it compare to the accuracy of traditional SOTA models?\\
RQ3. Ablation Research. How do the different parts contribute to improving the performance of the model?\\
RQ4. Performance under few-shot Settings.\\ 
RQ5. Uniformity and Alignment. How do the key metrics used to evaluate the embedding vectors change?\\
RQ6. Hyperparameters. How do different hyperparameters affect the performance of the model?\\
\subsection{Setup
	\label{sec:setups}}
\textbf{Environment}. Our experimental environment is set up on a Linux server, running Ubuntu 20.04, equipped with two Intel Xeon 4210r CPUs at 2.4Ghz (each CPU has 10 cores and 20 virtual cores), 128G memory, and 1 Nvidia RTX 3090 GPU (24G memory). The software environment is Python 3.8 and PyTorch 1.11.0 with CUDA 11.7.

\textbf{Hyperparameters}. SimCLF model is implement with two encoders. First one is TREX, a BERT based pre-trained model\citep{pei_TREX_nodate}, which has been pretrained for 10 epochs. For fine-tuning, the total number of epochs is set to 40, the temperature is set to 0.07, and the dimension of the output embedding is set to 768. We use the Adam optimizer with a learning rate of 0.00001 and set the maximum input length to 512. The model is trained with a weight decay of $1e$-$4$. And we use a batch size of 32. It is named SimCLF in this section. 

The second encoder is jTrans, a BERT based pre-trained model. It embeds control flow information of binary code into Transformer-based language models. For fine-tuning, the total number of epochs is set to 2, the temperature is set to 0.07, and the dimension of the output embedding to is set 512. The Adam optimizer is used with a learning rate of 0.000005 and the maximum input length is set to 512. We train the model with a weight decay of $1e$-$4$ and we use a batch size of 16. It is named jTrans-SimCLF in this section.

\textbf{Metrics}. In practice, functions that are compiled from the same source code are used as similar pairs. This saves a lot of manual labelling effort, and it is also a common practice in academia \citep{ding_asm2vec_2019}\citep{pei_TREX_nodate}. The cosine similarity function is utilized to calculate the distance between the embedding vectors of two functions. If the distance between a pair of functions is below a certain threshold, the two functions are classified as different. The robustness of our detection model is evaluated using several mechanism. The first one is AUC-ROC. Typically, the Receiver Operating Characteristic (ROC) curve \citep{ZHANG201696} of the model is plotted. The ROC curve has the false positive rate on the x-axis and the true positive rate on the y-axis. The area under the plotted ROC curve, known as the AUC (Area Under Curve) is used to assess the performance of the models. The AUC values range between 0.5 and 1, where a higher value indicates a better-performing model. 
MRR evaluates how well the system ranks the correct or relevant item at the top of the list. It considers the reciprocal rank of the first relevant item found for each query, and then calculates the average reciprocal rank across all queries. 
Recall@1 calculates the percentage of queries where the first result in the ranked list is a relevant item. It indicates how well the system is able to retrieve the most relevant item as the top recommendation.
\subsection{RQ1: Statistical Laws of Assembly Code}

\begin{figure}
		\centering
		\includegraphics[scale=0.4]{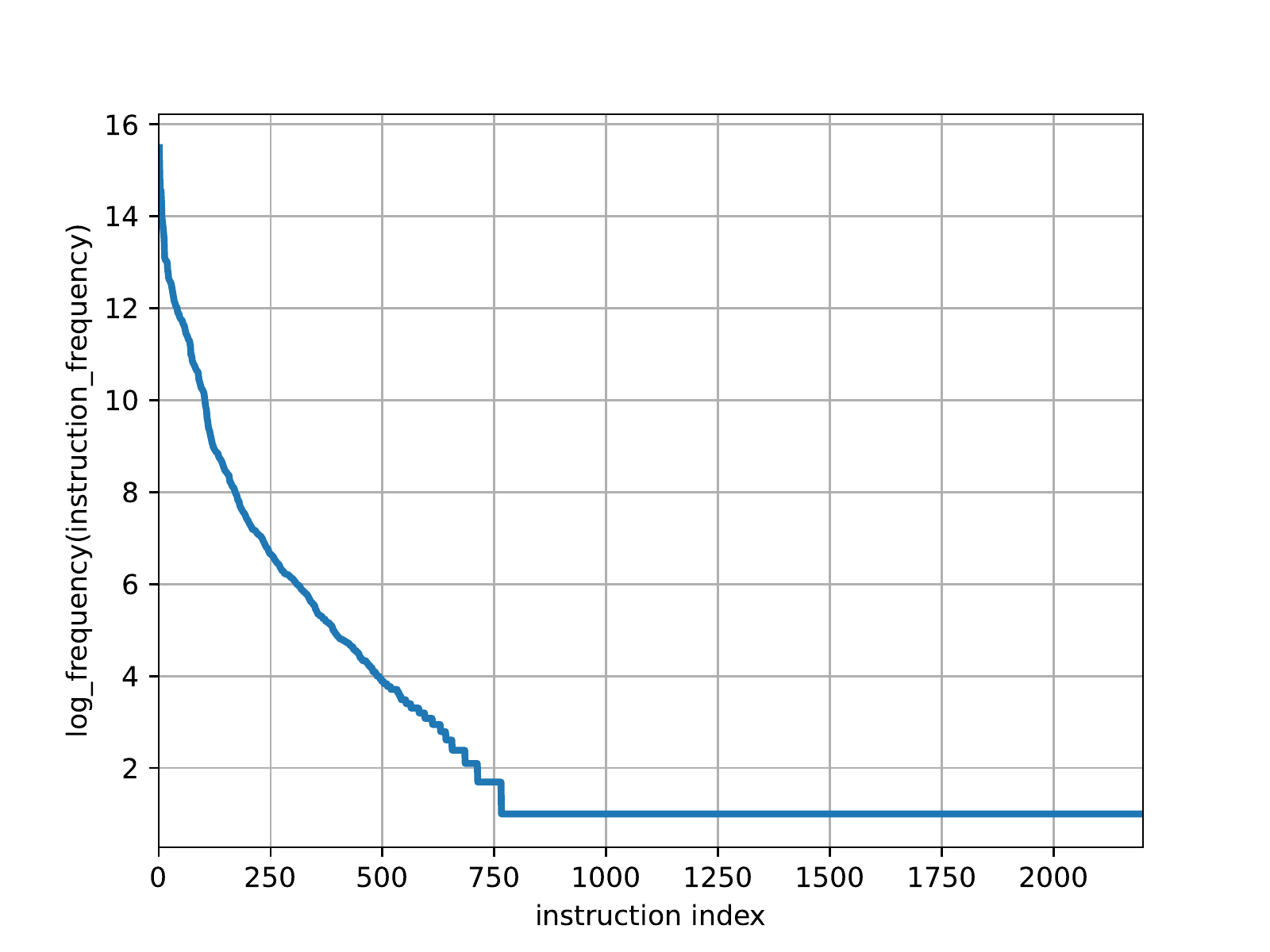}

	\caption{The frequency of the assembler word follows the Zipf distribution. The X-axis represents the rank of instructions in the corpus, while the Y-axis represents the logarithmic value of instruction frequencies. The graphic illustrates a long tail that has been rarely seen. }
	\label{fig:zipf}
\end{figure}
The initial question we address revolves around the distribution of assembly code frequencies and whether pre-trained models suffer from expression collapse issues. Despite the impressive results achieved by deep neural network-based encoders in binary code similarity detection, concerns may arise regarding the performance of these models due to the complexity and lack of interpretability associated with neural networks \citep{abs-2010-09470}. Several articles such as \citep{gao_representation_2019}\citep{li_sentence_2020} have highlighted the issue of expression collapse in pre-trained models within the field of NLP. This problem arises from the uneven distribution of word frequencies, wherein high-frequency words occur more frequently and therefore receive more gradient updates during the optimization process. Conversely, low-frequency words receive fewer updates, leading to a reduced impact on the embeddings. Consequently, the embeddings of high-frequency words tend to collapse towards a corner in the embedding space. A similar phenomenon has been observed in assembly language, as depicted in Figure \ref{fig:corner}. The embeddings of assembly tokens also seem to collapse into a corner. 

\begin{figure}
		\centering
		\includegraphics[scale=0.25]{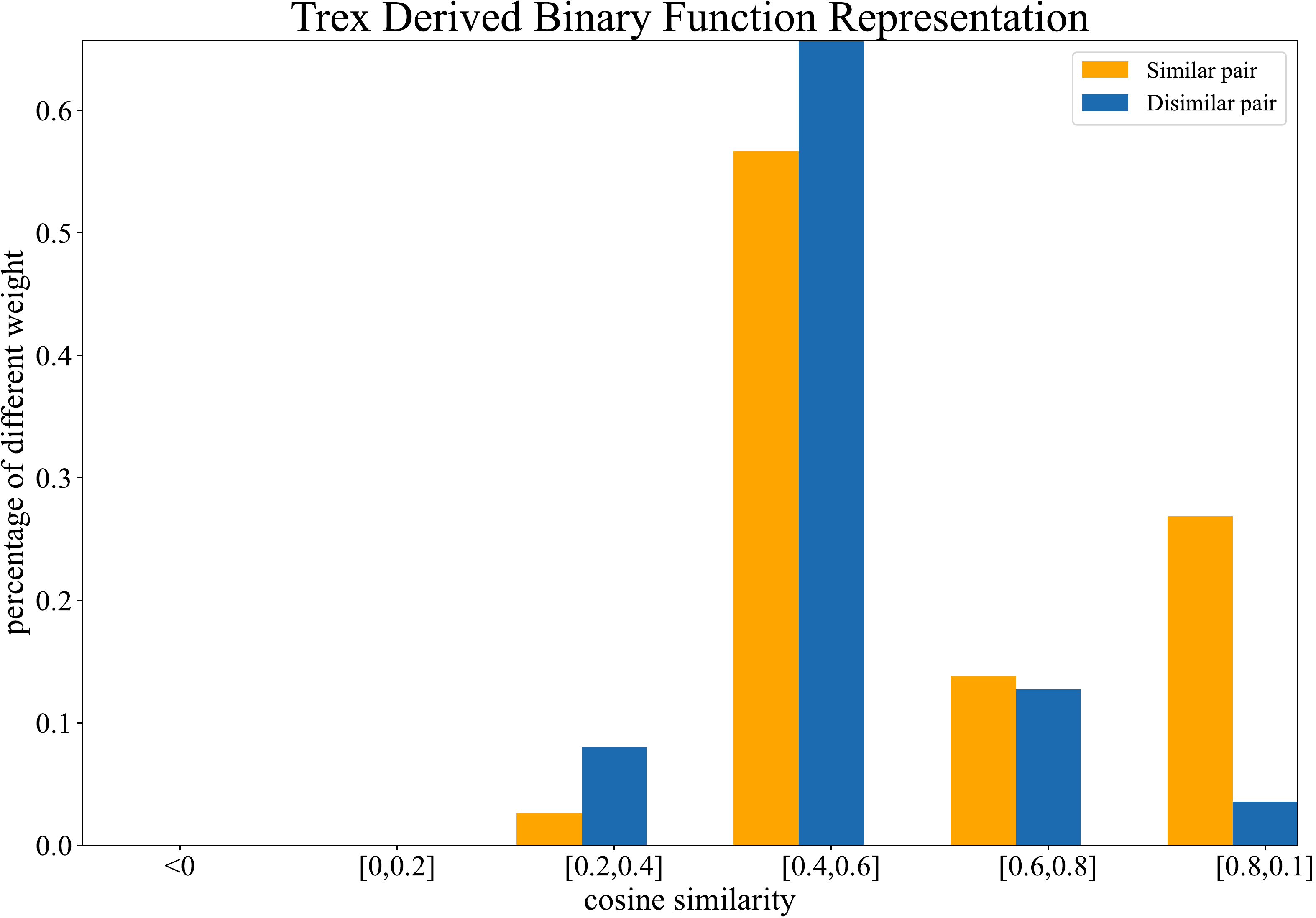}

	\caption{As shown in the figure, the yellow bar represents similar function pairs, and blue represents dissimilar function pairs. For the original TREX model, it has been observed that the majority of pairs of functions with the same meaning tend to receive a score of approximately 0.6. Similarly. Most pairs of functions with different meanings also tend to obtain a score around 0.6. Therefore, it can be concluded that the original TREX does not effectively distinguish between functions with different semantics. }
	\label{fig:before}
\end{figure}
So we have a conjecture that the distribution of the assembly code is also close to the Zipf distribution. To explore this hypothesis, we meticulously computed the frequency of each assembly token within the fine-tuning dataset documented in \citep{pei_TREX_nodate}, encompassing a substantial corpus of approximately 300 million tokens. The ensuing analysis culminated in the generation of Figure \ref{fig:zipf}, wherein the observed frequency distribution of the assembly code exhibits a notable resemblance to the characteristic pattern elicited by the Zipf distribution. We believe it is the reason that lead to the problem of expression collapse in the TREX model as shown in Figure \ref{fig:corner}. 

\begin{figure}[h]
		\centering
		\includegraphics[scale=0.25]{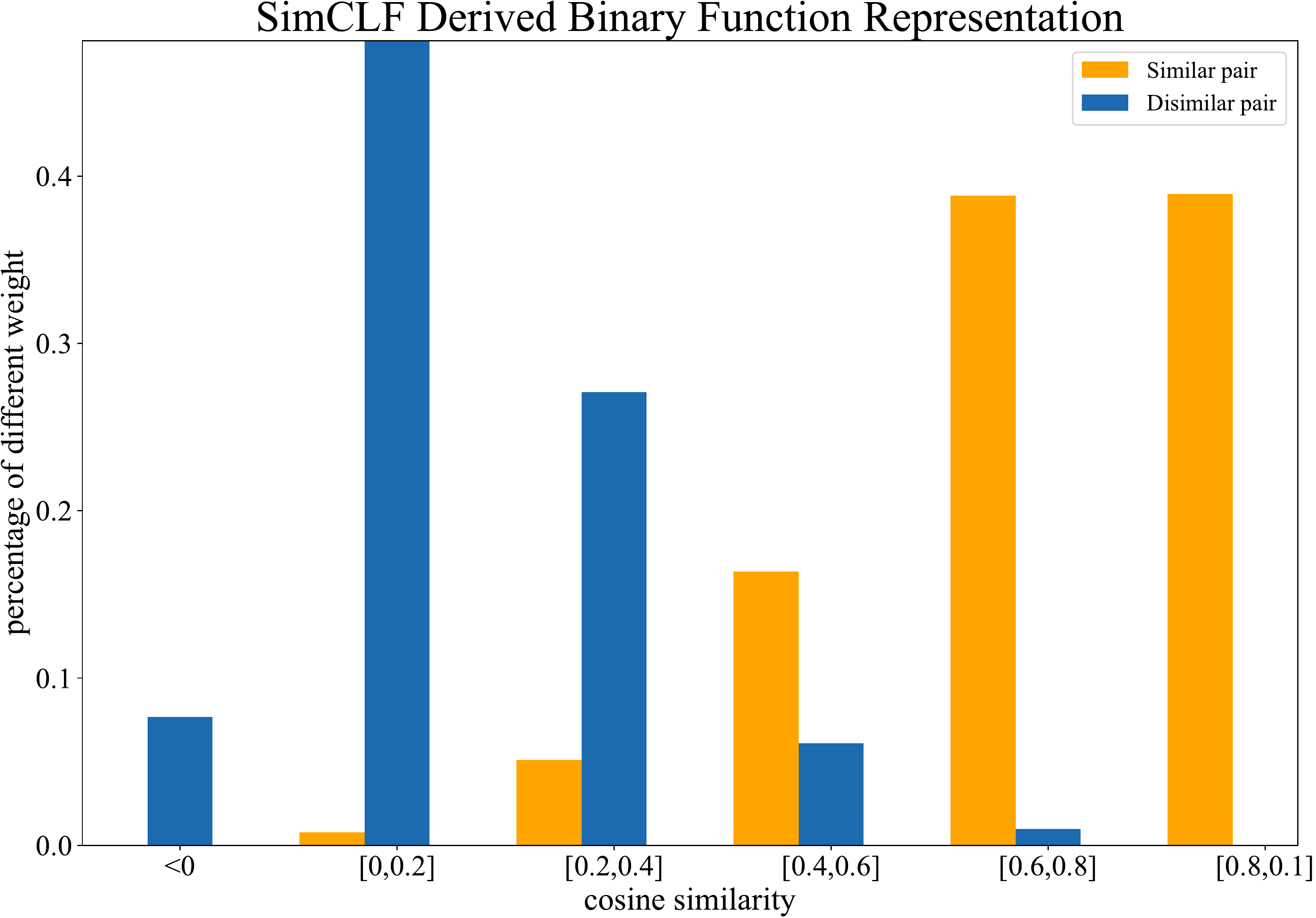}

	\caption{After fine-tuning the model, the obtained embeddings can be effectively employed to discern between similar and dissimilar functions with notable accuracy.}
	\label{fig:after}
\end{figure}

After observing the collapse of the embedded representation at the instruction level, it is reasonable to assume that the representation at the function level is also affected. The most commonly used approach to derive fixed-size sentence embedding is to average the BERT output layer \citep{reimers_sentence-bert_2019}. Prior to fine-tuning the TREX model, a direct adoption of average pooling to derive function representations by utilizing the token embeddings at the last layer leads to unsatisfactory outcomes. In other words, employing average pooling directly on the individual token embeddings at the final layer does not yield favorable results. Almost all function pairs achieved a similarity score of between $0.6$ and $1.0$, as illustrated in Figure \ref{fig:before}. It is notable that both similar and dissimilar function pairs exhibited cosine distances of approximately 0.6. However, after fine-tuning the model with SimCLF, we are able to successfully mitigate the anisotropy of the model \citep{ethayarajh_how_2019}. The utilization of average pooling to derive function embeddings enables effective differentiation between similar and dissimilar functions, as demonstrated in Figure \ref{fig:after}. Specifically, the cosine similarity of function pairs with identical semantics ranges from $[0.3,1]$, whereas dissimilar functions exhibit cosine similarities between $[-1,0.3]$. 
\subsection{RQ2:The Accuracy of The Model}
To evaluate the accuracy of SimCLF model, we conducted a comparative analysis with various SOTA models. The results of this evaluation demonstrated that SimCLF models consistently outperformed the other models, attaining the best overall results in terms of accuracy.
\subsubsection{Compared with SAFE model}

\begin{figure}[h]
	\begin{minipage}[t]{0.45\linewidth}
		\centering
		\includegraphics[scale=0.4]{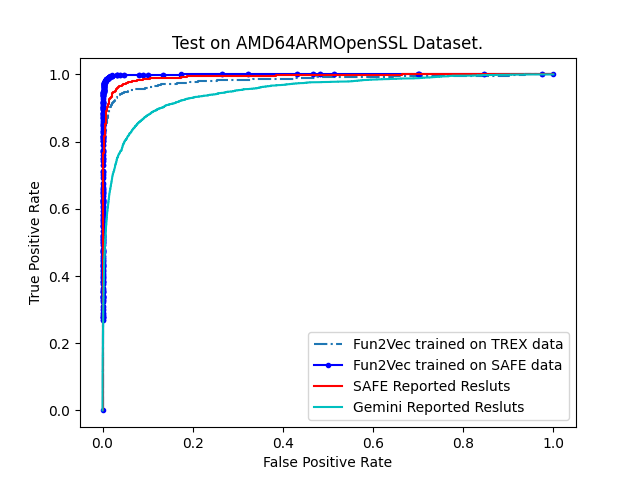}
		\centerline{(a) ROC curves on AMD64ARMOpenSSL Dataset.}
	\end{minipage}%
	\begin{minipage}[t]{0.45\linewidth}
		\centering
		\includegraphics[scale=0.4]{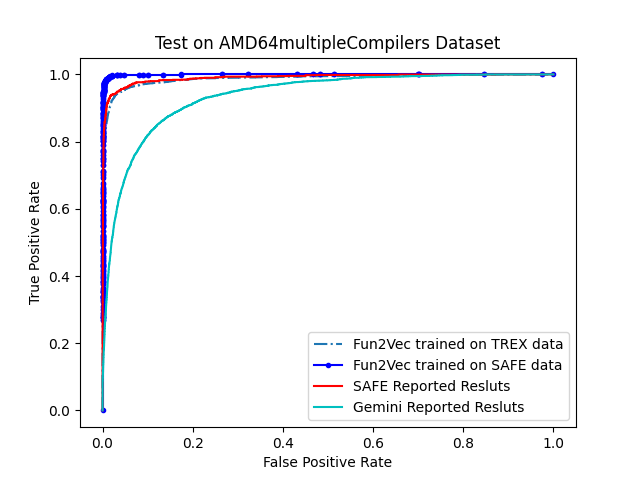}
		\centerline{(a) ROC curves on AMD64multipleCompilers Dataset.}
	\end{minipage}
	\caption{ROC curves were generated for the AMD64ARMOpenSSL Dataset and AMD64multipleCompilers Dataset, comparing the performance of different models. The curves of SimCLF are represented in blue, and they consistently outperform all the other models, as indicated by their superior performance across the evaluation metrics.}
	\label{figsafe}
\end{figure}
In the comparison between SimCLF and SAFE on the dataset published by SAFE \citep{massarelli_safe_2019}, we focus on two specific datasets. The first dataset is called AMD64ARMOpenSSL, which consists of 95,535 functions. All these functions have been compiled for both x86-64 and ARM architectures utilizing gcc-5.4 with four different optimization levels. The dataset used for comparison includes two open-source software repositories, namely OpenSSL-1.0.1f and OpenSSL-1.0.1u. Additionally, there is another dataset named AMD64multipleCompilers, which comprises a total of 452,598 functions. They come from the following open source software: binutils-2.30, ccv-0.7, coreutils-8.29, curl-7.61.0, gsl-2.5, libhttpd-2.0, openmpi-3.1.1, openssl-1.1.1-pre8, and valgrind-3.13.0. two experiments were conducted for each dataset. In the first experiment, our model was trained using the TREX data and then tested on the SAFE data. In the second experiment, SimCLF model was trained using both the AMD64ARMOpenSSL and AMD64multipleCompilers datasets. The results obtained from these experiments will be compared with the claims made by SAFE and Gemini \citep{xu_neural_2017}. 
The experiments were conducted with a total of two runs, and the obtained results were averaged to ensure accuracy and reliability. Figure \ref{figsafe} a illustrates the outcomes for the AMD64ARMOpenSSL case, revealing that SimCLF achieved a mean AUC testing score of 0.999. In comparison, SAFE obtained an AUC score of 0.992 for the same dataset. Similarly, in Figure \ref{figsafe} b, the results for the AMD64multipleCompilers case indicate that SimCLF attained a mean testing AUC score of 0.999, while SAFE achieved an AUC score of 0.990. Additionally, the evaluation result of Gemini \citep{xu_neural_2017} on the SAFE data \citep{massarelli_safe_2019} for further comparison purposes is also included. It is evident that SimCLF outperforms both the SAFE and Gemini models in terms of AUC scores on both datasets.\\
\begin{table}[htbp]
\caption{AUC scores on AMD64ARMOpenSSL and AMD64multipleCompilers.}

	\begin{tabular}{ccc}

		\toprule 
		Models&AMD64ARMOpenSSL&AMD64multipleCompilers\\
		\midrule 
		SimCLF trained on TREX& 0.982&0.987 \\
		SimCLF trained on SAFE &\textbf{0.999}&\textbf{0.999}	\\
		SAFE trained on SAFE &0.992&0.990	\\
		Gemini trained on SAFE&0.948&0.932\\
		\bottomrule[1.5pt]
	\end{tabular}
\end{table}

\subsubsection{Compared with TREX Model
	\label{TREXmodel}}
TREX is the state-of-the-art model proposed in \citep{pei_TREX_nodate}. It leverages a hierarchical Transformer architecture to acquire execution semantics from micro-traces in the pre-training phase. Subsequently, the authors employ InferSent \citep{conneau_supervised_2018} as a supervised embedding model to fine-tune the pre-trained TREX model. In this section, we conducted fine-tuning experiments on the TREX model using a shared dataset comprising 89,809 labelled function pairs. However, during the fine-tuning process, a subset of 22,453 positive examples from the shared dataset were selectively utilized for SimCLF model. Subsequently, the performance of SimCLF model were evaluated on the designated test dataset of the TREX model. The test dataset collects 13 popular open-source software projects. These projects include Binutils-2.34, Coreutils-8.32, Curl-7.71.1, Diffutils-3.7, Findutils-4.7.0, GMP-6.2.0, ImageMagick-7.0.10, Libmicrohttpd-0.9.71, LibTomCrypt-1.18.2, OpenSSL-1.0.1f and OpenSSL-1.0.1u, PuTTy-0.74, SQLite-3.34.0, and Zlib-1.2.11. As depicted in Table \ref{tab1}, function pairs were divided for each project (indicated in the first column) by employing five distinct types of partitioning strategies.
(1)ARCH: the function pairs have different architectures but the same optimizations without obfuscations (2nd column).
(2)OPT: the function pairs have different optimizations but the same architectures without obfuscations (3rd column). 
(3)OBF: the function pairs have different obfuscations with the same architectures (x64) and no optimization (4th column). 
(4)ARCH+OPT: the function pairs have both different architectures and optimizations without obfuscations (5th column). 
(5)ARCH+OPT+OBF: the function pairs can come from arbitrary architectures, optimizations, and obfuscations (6th column). A comparative analysis is conducted between SimCLF and TREX using identical test datasets, and the corresponding results are presented in Table \ref{tab1}. On average, SimCLF demonstrates a superior performance with an AUC score exceeding $0.984$. Notably, in the specific case of partition $3$, SimCLF achieves perfect performance, attaining an exceptional AUC score of $0.999$. Upon analyzing the results for partitions $1$, $2$, and $3$, it is observed that the performance of both SimCLF model and TREX model are quite similar. Due to the simplicity of the data in partitions $1$, $2$, and $3$, both models have reached the limits of accuracy. The disparity between the two models is minimal across these partitions.
However, as the complexity of the samples increases, SimCLF model demonstrates a notably superior performance compared to TREX. This observation suggests that SimCLF possesses enhanced analytical capabilities when dealing with challenging examples. Based on these findings, we can confidently conclude that SimCLF outperforms the original TREX model comprehensively in terms of overall performance.
\begin{table}[H]
	\caption{Results on function pairs across architectures, optimizations, and obfuscations.}
	\centering
	
	\resizebox{\linewidth}{!}{
		\begin{tabular}{ccc|cc|cc|cc|cc}
			\toprule[1.5pt]
			\multirow{2}{*}{Compiler}& \multicolumn{2}{c|}{ARCH} & \multicolumn{2}{c|}{OPT}&\multicolumn{2}{c|}{OBF}&\multicolumn{2}{c|}{ARCH+OPT}&\multicolumn{2}{c}{ARCH+OPT+OBF} \\
			&TREX& SimCLF& TREX& SimCLF& TREX&SimCLF& TREX&Fun2Vec& TREX&SimCLF\\
			\midrule
			Binutils& 	0.998 &0.998& 	0.993& 0.993& 	0.998& 	0.999&		0.975& 	0.996& 0.974& 0.994 \\
			Coreutils& 	0.998 &0.998& 	0.993& 0.994& 	0.998& 	0.999&		0.977& 	0.995& 0.973& 0.993 \\		
			Curl	& 	0.995 &0.995& 	0.989& 0.989& 	0.997& 	0.998&		0.972& 	0.990& 0.972& 0.992 \\		
			Diffutils&	0.998 &0.999& 	0.994& 0.994& 	0.998& 	0.999&		0.973&	0.991& 0.973& 0.992 \\
			Gmp& 		0.984 &0.984& 	0.982& 0.982& 	0.996& 	0.998&		0.972& 	0.990& 0.973& 0.990 \\
			Findutils& 	0.999 &0.999& 	0.991& 0.991& 	0.995&	0.997& 		0.975& 	0.993& 0.974& 0.993 \\
			ImageMagick & 0.984 &0.984& 0.978& 0.978& 	0.998&	0.999&		0.965&	0.987& 0.953& 0.980\\		
			Libmicrohttpd& 0.996 &0.996& 0.987&0.987& 	0.993&	0.999& 		0.965& 	0.986& 0.959& 0.978\\		
			LibTomCrypt& 0.983 &0.983& 	0.978& 0.995& 	0.998&	0.999& 		0.963& 	0.985& 0.949& 0.975 \\
			OpenSSL & 	0.996 &0.996& 	0.993& 0.993& 	0.998&	0.999& 		0.964& 	0.994& 0.953& 0.978 \\
			PuTTy & 	0.995 &0.995& 	0.979& 0.979& 	0.995&	0.997& 		0.957& 	0.977& 0.953& 0.977 \\
			SQLite & 	0.996 &0.996& 	0.979& 0.979& 	0.997&	0.999& 		0.957&	0.977& 0.953& 0.978 \\		
			Zlib & 	0.973 &0.974& 		0.973& 0.974& 	0.998&	0.999& 		0.957& 	0.976& 0.953& 0.978 \\		
			Average & \textbf{0.992} &\textbf{0.992}& 	0.986& \textbf{0.987}&	0.997&	\textbf{0.999}& 		0.967& 	\textbf{0.988}& 0.963& \textbf{0.984}\\
			\bottomrule[1.5pt]
			\label{tab1}
	\end{tabular}}
\end{table}
\subsubsection{Compared with jTrans Model}
jTrans \cite{wang2022jtrans} is a novel Transformer-based approach that aims to learn representations of binary code. It is the first solution to embed control flow information of binary code into Transformer-based language models, achieved through a jump-aware representation of the analyzed binaries and a newly-designed pre-training task. In our study, we conducted fine-tuning experiments on both the jTrans model and our SimCLF model, using a shared dataset BinaryCorp which is introduced by \cite{wang2022jtrans}. BinaryCorp comprises two datasets, namely BinaryCorp-3M and BinaryCorp-26M. BinaryCorp-3M includes 10,265 binary programs and approximately 3.6 million functions. BinaryCorp-26M includes 48,130 binary programs and approximately 26 million functions, surpassing GNUtils' 161,202 and Coreutils’ 76,174 functions by over 160 times and 339 times, respectively. The retrieval performance are evaluated using the two metrics MRR and Recall@1. In addition to jTrans, six other models are selected as baselines. They are Genius\cite{10.1145/2976749.2978370}, Gemini \cite{xu_neural_2017}, SAFE \cite{massarelli_safe_2019}, Asm2Vec \cite{ding_asm2vec_2019}, GraphEmb \cite{massarelli_investigating_2019}, 
 OrderMatters\cite{yu_order_2020}. On each dataset, two function pool sizes 32 and 10,000 are used. In this test, our model is illustrated as jTrans-SimCLF. jTrans-zero is the model without finetured.
 \begin{table}[H]
	\caption{Results of different binary similarity detection methods on BinaryCorp-3M (Poolsize=32).}
	\centering
	
	\resizebox{\linewidth}{!}{

		\begin{tabular}{cccccccc|ccccccc}
  \toprule[1.5pt]

&\multicolumn{7}{c|}{MRR} &\multicolumn{7}{c}{Recall@1} \\
		\midrule 
Models& O0,O3& O1,O3& O2,O3& O0,Os& O1,Os& O2,Os& Average& O0,O3& O1,O3& O2,O3& O0,Os& O1,Os& O2,Os& Average\\
		\midrule 
Gemini&0.388&0.580&0.750&0.455&0.546&0.614&0.556&0.238&0.457&0.669&0.302&0.414&0.450&0.422\\
SAFE&0.826&0.917&0.958&0.854&0.927&0.927&0.902&0.729&0.869&0.933&0.766&0.879&0.880&0.843\\
Asm2Vec&0.479&0.878&0.961&0.536&0.855&0.900&0.768&0.351&0.828&0.942&0.408&0.796&0.863&0.701\\
GraphEmb&0.602&0.694&0.750&0.632&0.674&0.675&0.671&0.485&0.600&0.678&0.521&0.581&0.584&0.575\\
OrderMatters-online&0.542&0.740&0.869&0.638&0.702&0.682&0.695&0.414&0.647&0.822&0.515&0.611&0.593&0.591\\
OrderMatters&0.601&0.838&0.933&0.701&0.812&0.800&0.777&0.450&0.763&0.905&0.566&0.724&0.715&0.687\\
Genius&0.377&0.587&0.868&0.437&0.600&0.627&0.583&0.243&0.479&0.830&0.298&0.490&0.526&0.478\\
		\midrule 
jTrans-Zero&0.594&0.841&0.962&0.649&0.850&0.891&0.797&0.499&0.803&0.945&0.566&0.808&0.853&0.746\\
jTrans&0.947&0.976&0.985&0.956&0.979&0.977&0.970&0.913&0.960&0.974&0.927&0.964&0.961&0.949\\
		\midrule 
jTran-SimCLF&\textbf{0.961}&\textbf{0.978}&\textbf{0.986}&\textbf{0.969}&\textbf{0.981}&\textbf{0.980}&\textbf{0.976}
&\textbf{0.934}&\textbf{0.963}&\textbf{0.975}&\textbf{0.947}&\textbf{0.967}&\textbf{0.964}&\textbf{0.958}\\
		\bottomrule[1.5pt]

		\label{tabjtran3m32}
 
	\end{tabular}}
\end{table}
\begin{table}[H]
	\caption{Results of different binary similarity detection methods on BinaryCorp-3M (Poolsize=10000)}
	\centering
	
	\resizebox{\linewidth}{!}{

		\begin{tabular}{cccccccc|ccccccc}
  \toprule[1.5pt]

&\multicolumn{7}{c|}{MRR} &\multicolumn{7}{c}{Recall@1} \\
		\midrule 
Models& O0,O3& O1,O3& O2,O3& O0,Os& O1,Os& O2,Os& Average& O0,O3& O1,O3& O2,O3& O0,Os& O1,Os& O2,Os& Average\\
		\midrule 
Gemini&0.037&0.161&0.416&0.049&0.133&0.195&0.165&0.024&0.122&0.367&0.030&0.099&0.151&0.132\\
SAFE&0.127&0.345&0.643&0.147&0.321&0.377&0.320&0.068&0.247&0.575&0.079&0.221&0.283&0.246\\
Asm2Vec&0.072&0.449&0.669&0.083&0.409&0.510&0.366&0.046&0.367&0.589&0.052&0.332&0.426&0.302\\
GraphEmb&0.087&0.217&0.486&0.110&0.195&0.222&0.219&0.050&0.154&0.447&0.063&0.135&0.166&0.169\\
OrderMatters&0.062&0.319&0.600&0.075&0.260&0.233&0.263&0.040&0.248&0.535&0.040&0.178&0.158&0.200\\
Genius&0.041&0.193&0.596&0.049&0.186&0.224&0.214&0.028&0.153&0.538&0.032&0.146&0.180&0.179\\
		\midrule 
jTrans-Zero&0.137&0.490&0.693&0.182&0.472&0.510&0.414&0.088&0.412&0.622&0.122&0.393&0.430&0.340\\
jTrans&0.475&0.663&0.731&0.539&0.665&0.664&0.623&0.376&0.580&0.661&0.443&0.586&0.585&0.571\\
		\midrule 
jTran-SimCLF&\textbf{0.557}&\textbf{0.695}&\textbf{0.741}&\textbf{0.609}&\textbf{0.691}&\textbf{0.691}&\textbf{0.664} &\textbf{0.464}&\textbf{0.619}&\textbf{0.674}&\textbf{0.520}&\textbf{0.616}&\textbf{0.616}&\textbf{0.585}\\
		\bottomrule[1.5pt]

		\label{tabjtran3m10000}
 
	\end{tabular}}
\end{table}

The results of the experiments conducted on BinaryCorp-3M are presented in Table \ref{tabjtran3m32} and Table \ref{tabjtran3m10000}. For poolsize=32 (Table \ref{tabjtran3m32}), jTrans-SimCLF demonstrates superior performance compared to its closest baseline competitor, with an average improvement of 0.06 for the MRR metric and over 0.09 for the recall@1 metric. When considering poolsize=10000, jTrans-SimCLF outperforms all baselines by significant margins. On average, jTrans-SimCLF surpasses its closest competitor, jTrans, by 0.041 for the MRR metric and over 0.014 for the recall@1 metric. Notably, jTrans-SimCLF achieves these results with just two training epochs, while jTrans was trained for ten epochs.

The results of our experiments conducted on BinaryCorp-26M are presented in Table \ref{tabjtran26m32} and Table \ref{tabjtran26m10000}. jTrans-SimCLF demonstrates superior performance compared to all baselines across all test scenarios. Specifically, for poolsize=32, jTrans-SimCLF outperforms jTrans by an average improvement of 0.02 for the MRR metric and over 0.01 for the recall@1 metric. Furthermore, when considering poolsize=10000, jTrans-SimCLF surpasses all baselines by significant margins. On average, jTrans-SimCLF outperforms its closest competitor, jTrans, by 0.018 for the MRR metric and over 0.021 for the recall@1 metric. These findings highlight the robust performance of jTrans-SimCLF in comparison to alternative approaches. Notably, jTrans-SimCLF achieves these results with just two training epochs and half of the dataset, while jTrans was trained for ten epochs on the complete dataset.

\begin{table}[h]
	\caption{Results of different binary similarity detection methods on BinaryCorp-26M (Poolsize=32).}
	\centering
	
	\resizebox{\linewidth}{!}{

		\begin{tabular}{cccccccc|ccccccc}
  \toprule[1.5pt]

&\multicolumn{7}{c|}{MRR} &\multicolumn{7}{c}{Recall@1} \\
		\midrule 
Models& O0,O3& O1,O3& O2,O3& O0,Os& O1,Os& O2,Os& Average& O0,O3& O1,O3& O2,O3& O0,Os& O1,Os& O2,Os& Average\\
		\midrule 
Gemini&0.402&0.643&0.835&0.469&0.564&0.628&0.590&0.263&0.528&0.768&0.322&0.441&0.518&0.473\\
SAFE&0.856&0.940&0.970&0.874&0.935&0.934&0.918&0.770&0.902&0.951&0.795&0.891&0.891&0.867\\
Asm2Vec&0.439&0.847&0.958&0.490&0.788&0.849&0.729&0.314& 0.789&0.940&0.362&0.716&0.800&0.654\\
GraphEmb&0.583&0.681&0.741&0.610&0.637&0.639&0.649&0.465& 0.586&0.667&0.499&0.541&0.543&0.550\\
OrderMatters&0.572&0.820&0.932&0.630&0.692&0.771&0.729& 0.417&0.740&0.903&0.481&0.692&0.677&0.652\\

		\midrule 
jTrans-Zero&0.632&0.871&0.973&0.687&0.890&0.891&0.824& 0.539&0.838&0.961&0.602&0.854&0.853&0.775\\
jTrans&0.964&0.983&0.989&0.969&0.980&0.980&0.978&0.941& 0.970&0.981&0.949&0.964&0.964&0.962\\
		\midrule 
jTran-SimCLF&\textbf{0.967}&\textbf{0.983}&\textbf{0.991}&\textbf{0.972}&\textbf{0.981}&\textbf{0.982}&\textbf{0.980}
&\textbf{0.946}&\textbf{0.971}&\textbf{0.981}&\textbf{0.953}&\textbf{0.965}&\textbf{0.965}&\textbf{0.963}\\

		\bottomrule[1.5pt]

		\label{tabjtran26m32}
 
	\end{tabular}}
\end{table}

\begin{table}[h]
	\caption{Results of different binary similarity detection methods on BinaryCorp-26M (Poolsize=10000)}
	\centering
	
	\resizebox{\linewidth}{!}{

		\begin{tabular}{cccccccc|ccccccc}
  \toprule[1.5pt]

&\multicolumn{7}{c|}{MRR} &\multicolumn{7}{c}{Recall@1} \\
		\midrule 
Models& O0,O3& O1,O3& O2,O3& O0,Os& O1,Os& O2,Os& Average& O0,O3& O1,O3& O2,O3& O0,Os& O1,Os& O2,Os& Average\\
		\midrule 
Gemini&0.072&0.189&0.474&0.069&0.147&0.202&0.192&0.058& 0.148&0.420&0.051&0.115&0.162&0.159\\
SAFE&0.198&0.415&0.696&0.197&0.377&0.431&0.386&0.135&0.314& 0.634&0.127&0.279&0.343&0.305\\
Asm2Vec&0.118&0.443&0.703&0.107&0.369&0.480&0.370&0.099& 0.376&0.638&0.086&0.307&0.413&0.320\\
GraphEmb&0.116&0.228&0.498&0.133&0.198&0.224&0.233&0.080 &0.171&0.465&0.090&0.145&0.175&0.188\\
OrderMatters&0.113&0.292&0.682&0.118&0.256&0.295&0.292& 0.094&0.222&0.622&0.093&0.195&0.236&0.244\\

		\midrule 
jTrans-Zero&0.215&0.570&0.759&0.233&0.571&0.563&0.485& 0.167&0.503&0.701&0.175&0.507&0.500&0.426\\
jTrans &0.584 &0.734& 0.792 &0.627 &0.709& 0.710 &0.693& 0.499 &0.668& 0.736& 0.550& 0.648 &0.648&0.625\\
		\midrule 
jTran-SimCLF&\textbf{0.619}&\textbf{0.740}&\textbf{0.801}&\textbf{0.665}&\textbf{0.718}&\textbf{0.720}&\textbf{0.711} &\textbf{0.538}&\textbf{0.691}&\textbf{0.742}&\textbf{0.593}&\textbf{0.657}&\textbf{0.657}&\textbf{0.646}\\
		\bottomrule[1.5pt]

		\label{tabjtran26m10000}
 
	\end{tabular}}
\end{table}
\subsubsection{Vulnerability Search}
\begin{table}[h]
	\caption{Basic information of the Vulnerability dataset}
	\centering
	
	\begin{tabular}{ccccc}
		\toprule 
		\#&Alias/Method&CVE/stats&Function name& Number\\
		\midrule 
		1&Heartbleed&cve-2014-0160&tls1\_{}process\_{}heartbeat&8 \\
		2&Venom 	&cve-2015-3456&fdctrl\_{}handle\_{}drive\_{}specification\_{}command&6\\
		3&WS-snmp	&cve-2011-0444&snmp\_{}usm\_{}password\_{}to\_{}key\_{}sha1&7\\
		4&wget		&cve-2014-4877&ftp\_{}syst&5\\
		\bottomrule[1.5pt]
		
	\end{tabular}
	\label{table:exploit}
\end{table}
In order to further assess the performance of SimCLF, SimCLF is applied to a real-world application scenario. Specifically, SimCLF is employed to conduct a search for vulnerable functions within a publicly available dataset \citep{david_statistical_nodate}. The dataset utilized in this evaluation consists of multiple vulnerable binaries that were compiled using $11$ different compilers belonging to the clang, gcc, and icc families\citep{massarelli_safe_2019}. A subset functions that consisting of functions with a length of less than $512$ are selected. This subset contains a total of $2,220$ functions, out of which $26$ functions are identified as containing exploits. The dataset encompasses four distinct types of vulnerabilities, which are documented in Table \ref{table:exploit}. To process the dataset, we disassembled the functions in dataset using the $objdump$ tool and transformed the target functions into embeddings. Subsequently, these embeddings are compared with the functions known to contain exploits. In this vulnerability search experiment, SimCLF performed almost perfectly. It successfully identified all $26$ vulnerable functions in the top-k results for each query. Specifically, when considering the case where the number of functions containing vulnerabilities is denoted as $k$, SimCLF consistently discovered all $k$ vulnerable functions within the first $k$ results for every query.
Comparatively, the SAFE model, as documented in the paper by \citep{massarelli_safe_2019}, achieved a recall rate of $84$\% for $k$ = $10$. Meanwhile, the Gemini model achieved a lower recall rate of $55$\% for the same k value. This suggests that SimCLF outperformed both SAFE and Gemini in terms of recall for identifying vulnerable functions.
\subsection{RQ3:Ablation Analysis}
In this section, a quantitative analysis is employed to assess the influence of various components within the SimCLF model. With the exception of Subsection \ref{subsection_aumentation}, all reported results in this section are derived from the following standardized settings. All models were trained utilizing the parameters outlined in Section \ref{sec:setups}, and subsequently evaluated on a total of 13 distinct datasets as described in Section \ref{TREXmodel}. By maintaining consistent employment of these settings throughout all experiments, the precise impact of individual elements within the SimCLF framework can be accurately quantified.
\subsubsection{Projection Head}
\begin{figure}[h]
	\begin{minipage}[b]{0.5\linewidth}
		\subfloat[][model without head]{
			\centering
			\includegraphics[scale=0.3]{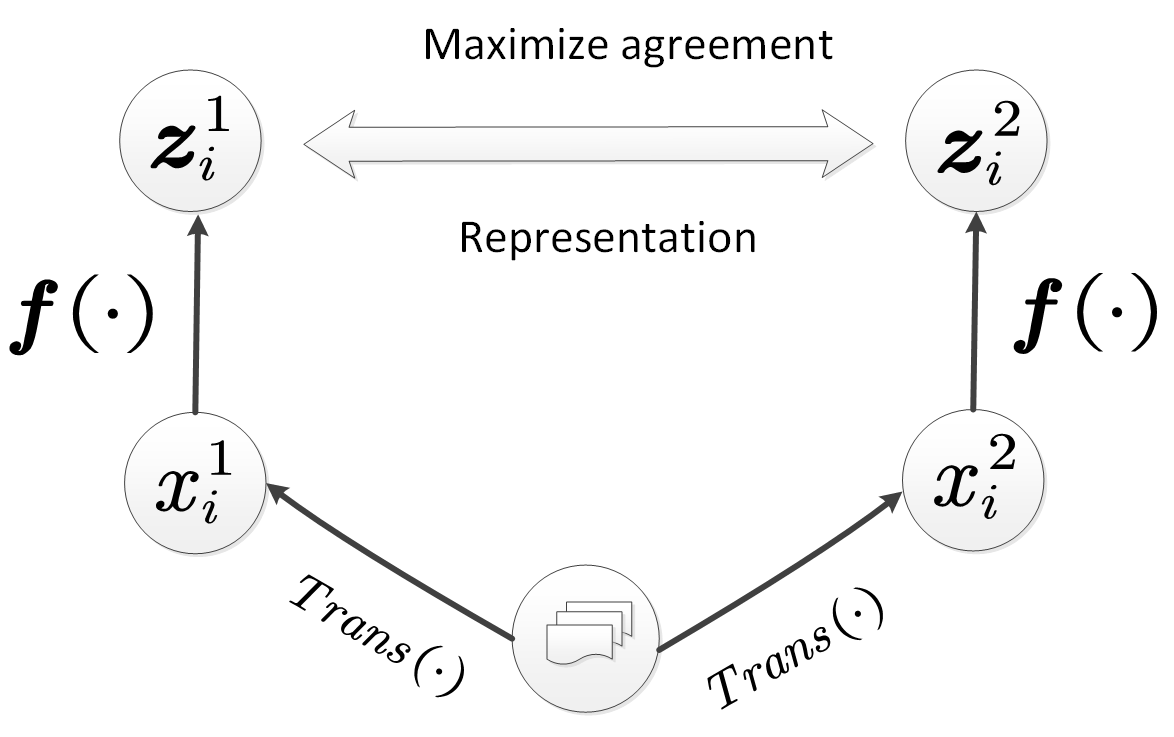}
		}
	\end{minipage}%
	\hfill
	\begin{minipage}[b]{0.5\linewidth}
		\subfloat[][model with head]{
			\centering
			\includegraphics[scale=0.34]{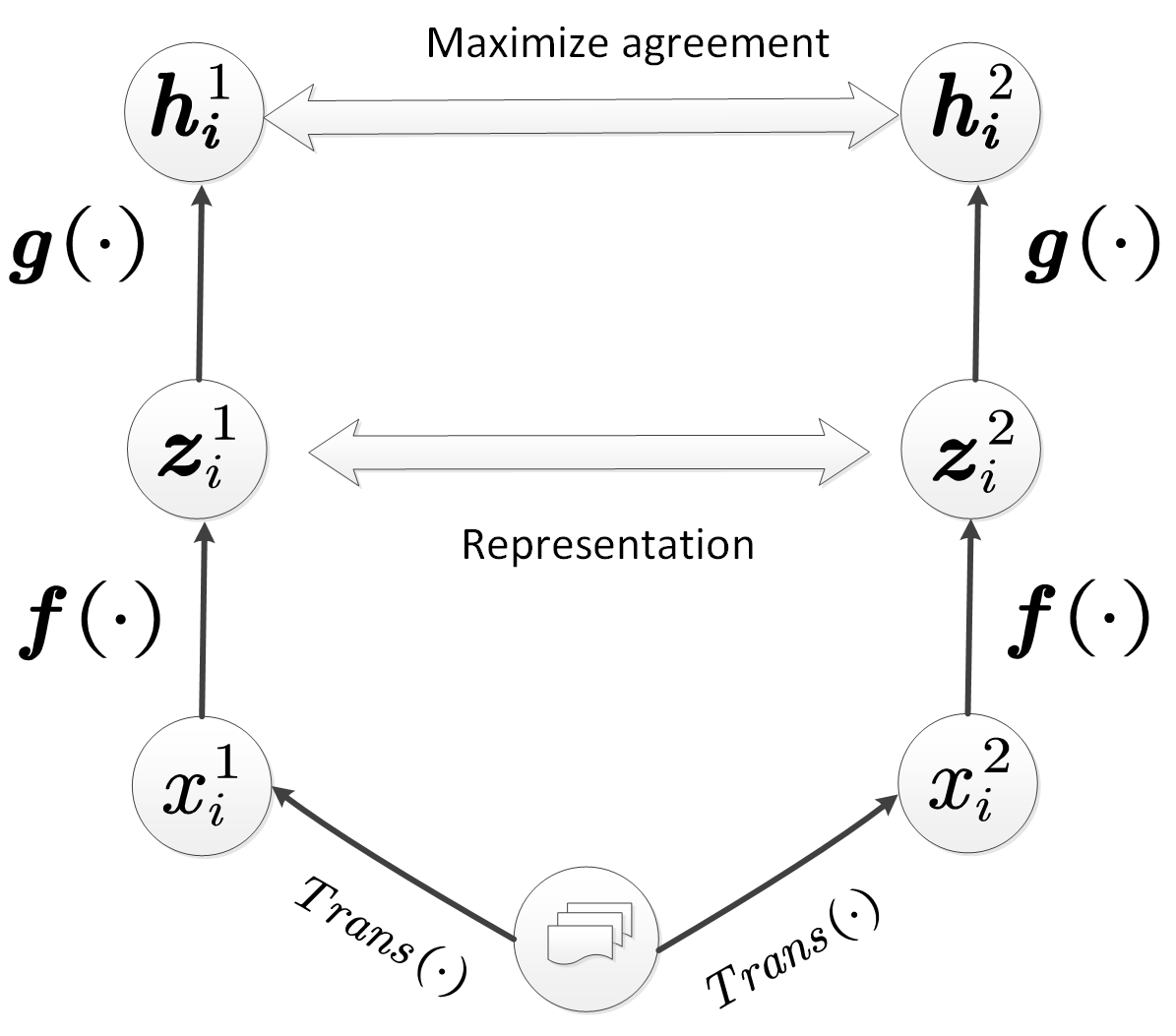}
		}

	\end{minipage}
	
	\caption{Schematic diagram of the structure of the models with and without project head.}
	\label{figwithhead}
\end{figure}

Several articles have pointed out that project head can improve the performance of the model before it \citep{DBLP:journals/corr/abs-2003-04297}\citep{chen_simple_2020}\citep{DBLP:journals/corr/abs-2006-07733}\citep{DBLP:journals/corr/abs-2006-10029}. The influences of including a projection head $g(h)$ are studied in this section. $g(h)$ maps representations to the space where the contrastive loss is applied after the training process. However, after the finetuning phase, the projection head $g(h)$ is discarded and not used further. Figure \ref{figwithhead} shows the difference between the model with and without the projection head. The embedding $z$ is used as the representation for downstream tasks.

\begin{figure}[h]
	\centering 
	\includegraphics[scale=0.3]{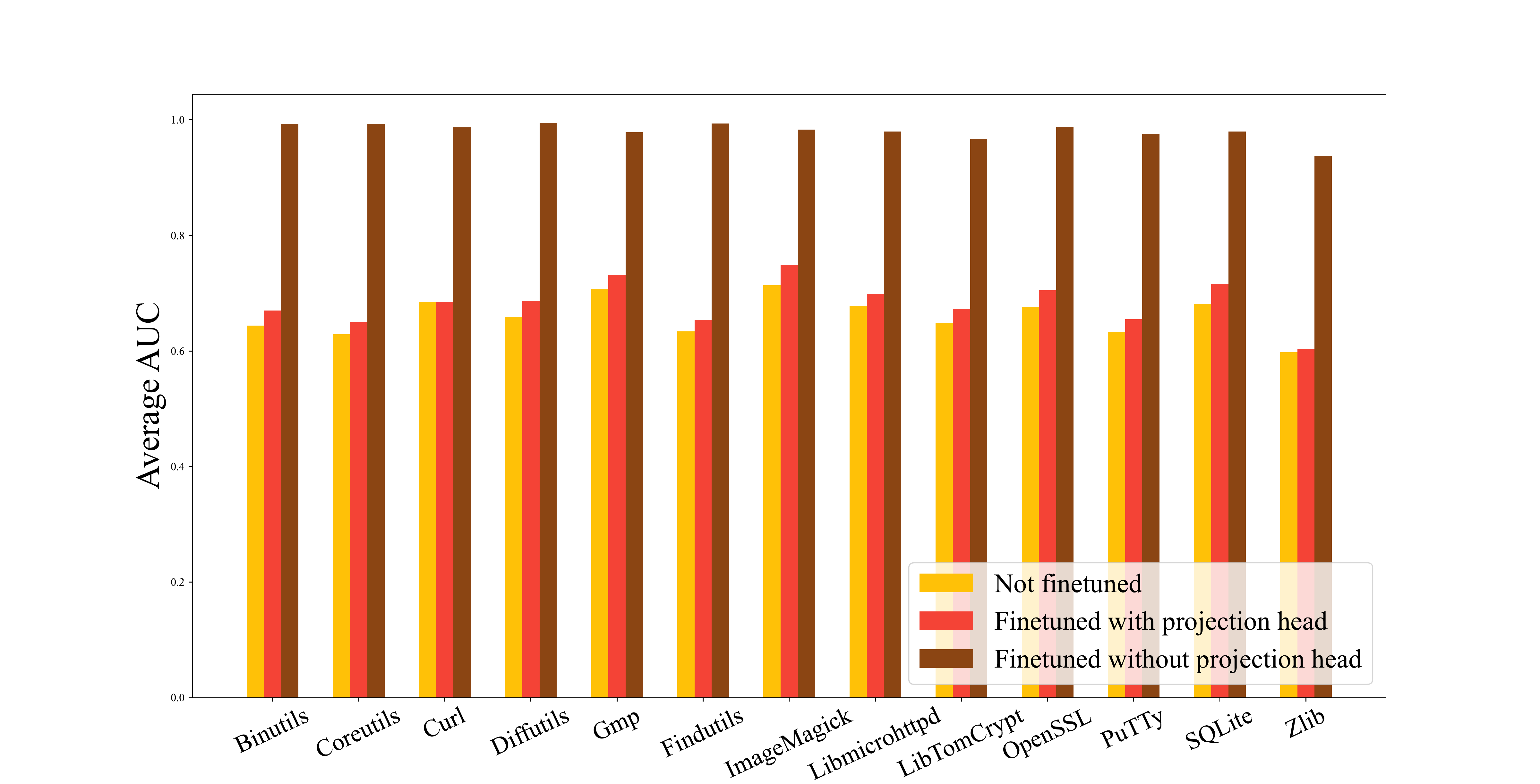} 
	\caption{Comparison of AUC scores on test set between models pre-trained with and without projection head.} 
	\label{figprojectionhead}
\end{figure}

Numerous scholarly articles \citep{DBLP:journals/corr/abs-2003-04297} \citep{chen_simple_2020} \citep{DBLP:journals/corr/abs-2006-07733}\citep{DBLP:journals/corr/abs-2006-10029} have employed the contrastive learning model as a prominent approach. While the function of the projection head, denoted as $g(h)$, has been widely acknowledged and demonstrated in various studies, such as SimCLR \citep{chen_simple_2020}, the precise reasons for its effectiveness remain subjects of ongoing debate. However, it is noteworthy that in SimCLF, incorporating the projection head did not yield desirable outcomes. Figure \ref{figprojectionhead} shows that the model’s AUC score increases slightly (on average 3\%) when the model is trained with projection head $g(h)$ compared to the model without fine-tuning. Nevertheless, it is worth noting that the AUC score of the model with the projection head remains considerably lower than that of the model without the projection head. A conjecture put forward by SimCLR \cite{chen_simple_2020} suggests that the significance of utilizing the representation prior to the nonlinear projection lies in mitigating the loss of information induced by the contrastive model. Additionally, another plausible explanation is that \citep{chen_simple_2020} incorporates a vast amount of data, wherein some information may be compromised when employing the contrastive model with the projection head. However, in our experiments the amount of data is relatively small (tens of thousands). In the process of fine-tuning, the BERT-based encoder is used. Compared to ResNet-50 \citep{chen_simple_2020}, BERT has a better encoding capability. The information lost due to contrastive learning is relatively small and could therefore be ignored.
\subsubsection{Pretraining Effectiveness.}
\begin{figure}[h]
	\centering 
	\includegraphics[scale=0.3]{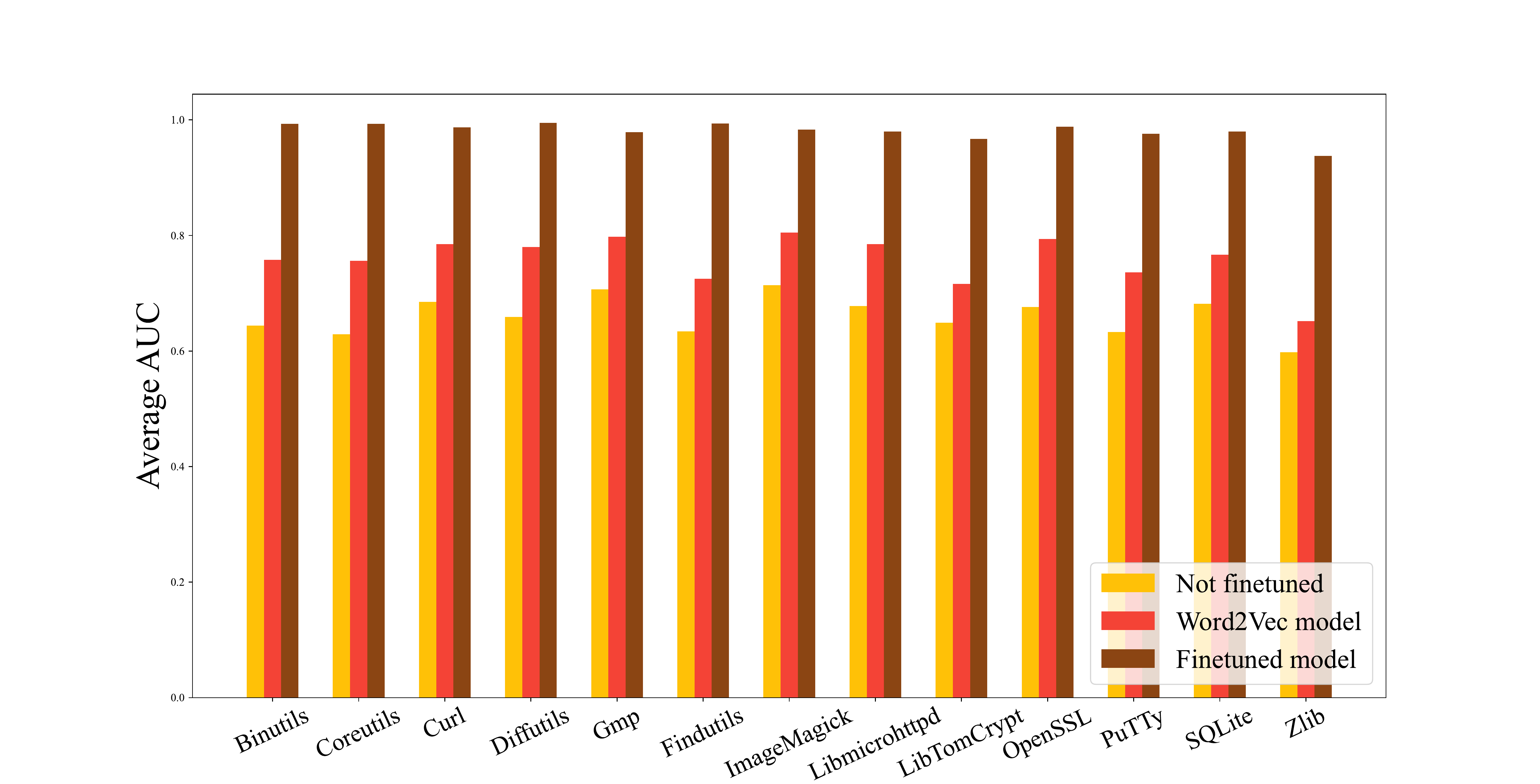} 
	\caption{Comparison of testing AUC scores between models pre-trained with and without the pre-training modules.} 
	\label{figword2vec}
\end{figure}

In order to delve deeper into the impact of the pre-training modules, a comparative analysis is conducted by removing the pre-training modules. Specifically, when the attention module is removed, SimCLF model reverts to a Word2Vec model. The AUC score of the model on various datasets is presented in Figure \ref{figword2vec}. The results reveal an average increase of 10\% in the AUC score. This finding underscores the crucial role played by the pre-training module in enhancing model performance and highlights its significance in the overall architecture.

\subsubsection{Only Finetuning The Projection Head}
\begin{figure}[h]
	\centering 
	\includegraphics[scale=0.25]{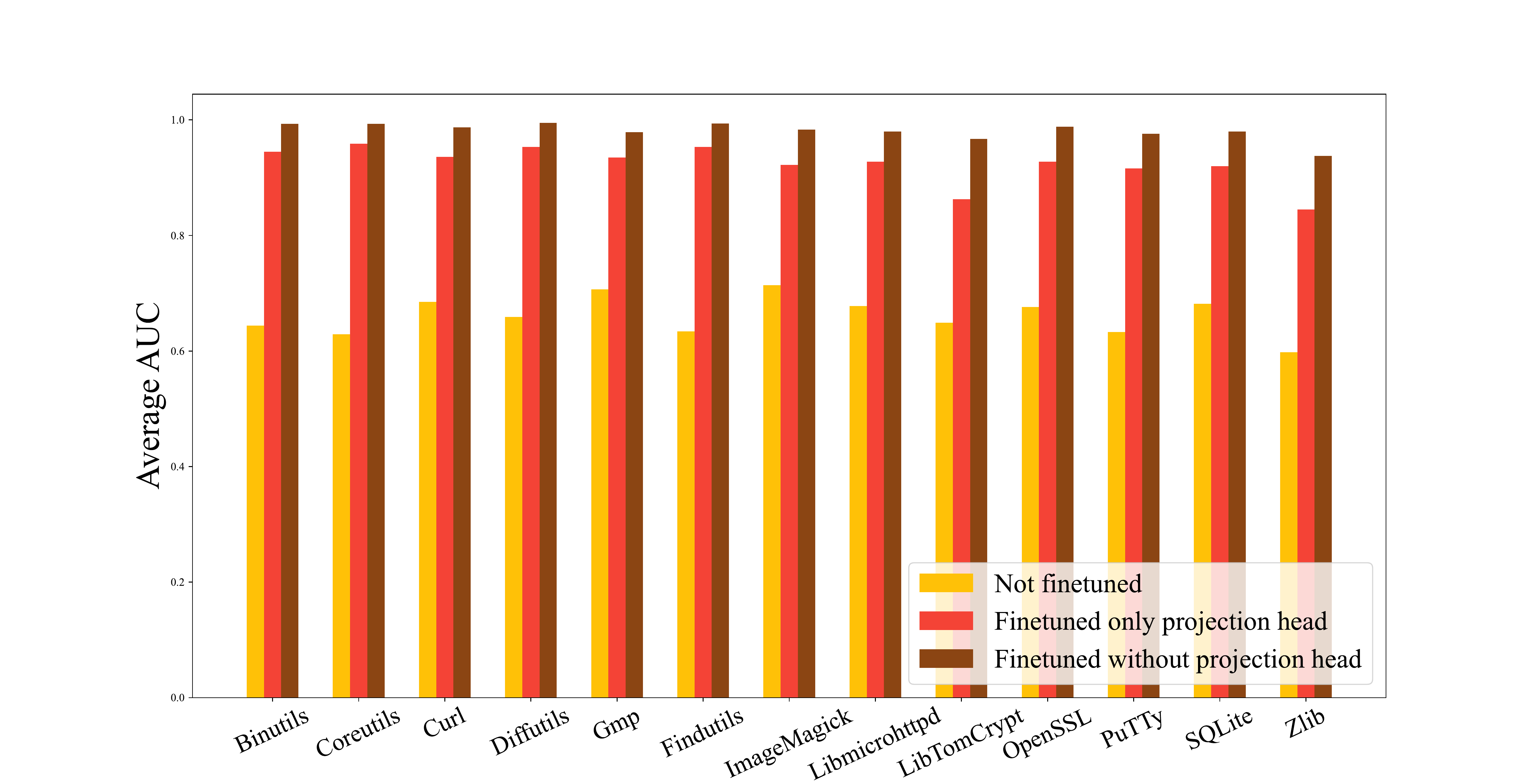} 
	\caption{Comparison of testing AUC scores between models only finetuning the projection head and SimCLF.} 
	\label{figonlyhead}
\end{figure}

Although the model with the addition of the projection head could not yield good results, we also tried different network structures to find the best network. A full connection layer was added to the TREX model, and when finetuning, the pre-trained model is fixed. Only the project heads are finetuned and the projection heads are kept. Figure \ref{figonlyhead} shows that when only the projection heads are finetuned, the AUC value decreases slightly, about 0.05-0.08.

\subsubsection{The data augmentation
\label{subsection_aumentation}}
\begin{table}[h]
	\caption{Downstream performance under different richness of augmentations}
	\begin{tabular}{c|ccccc|cc|cc}
		\toprule 
		 &\multicolumn{5}{c|}{Transformations}&\multicolumn{2}{c|}{Poolsize=32}&\multicolumn{2}{c}{Poolsize=10000}\\
 Model&o1&o2&o0&o3&os&MRR&Recall&MRR&Recall@1\\
		\midrule 
 jTrans-SimCLF& \checkmark & \checkmark & \checkmark & \checkmark & \checkmark & 0.980&0.963& 0.664&0.585 \\
 jTrans-SimCLF& \checkmark & \checkmark & \checkmark & \checkmark &      & 0.971&0.952& 0.656&0.577 \\
 jTrans-SimCLF& \checkmark & \checkmark & \checkmark &      &      & 0.963&0.9428& 0.638&0.559 \\
 jTrans-SimCLF& \checkmark & \checkmark &       &      &      & 0.951&0.928& 0.614&0.533 \\

		\bottomrule[1.5pt]
	\end{tabular}
 \label{table_augmentation}
\end{table}
To gain insight into the internal mechanisms of SimCLF, an empirical investigation was conducted to examine the correlation between data augmentation and downstream performance. All tests were performed on the jTrans-SimCLF model. The models were trained utilizing the parameters outlined in Section \ref{sec:setups} and subsequently evaluated on the BinaryCorp-3M dataset.

To evaluate the efficacy of the jTrans-SimCLF model, all five types of transformations were amalgamated, and one of the combined transformations was subsequently eliminated in a stepwise manner from five to two, resulting in four experiments being conducted on the dataset. As shown in Table \ref{table_augmentation}, it was observed that there was a monotonic decline in downstream performance as the number of transformations decreased. Therefore, it can be conclude that the use of data augmentation techniques is beneficial for the success of the model.

\subsection{RQ4:Performance under Few-shot Settings}
To validate the performance of SimCLF under the data scarcity scenarios, the experiments under the few-shot setting are performed. The number of unlabelled function pairs are limited to $2$, $8$, $32$, $128$, $512$ and $2048$ respectively. Then the performance of these models are compared with the model trained on the entire dataset. The average AUC values obtained from these comparisons are presented below:
\begin{table}[h]
	\caption{Performance under the Few-shot Settings}
	\begin{tabular}{ccccccccc}
		\toprule 
		Number&0&2&8&32&128&512&2048&All(22453)\\
		\midrule 
		AUC score& 0.679&0.719&0.737&0.777&0.859&0.929&0.962&0.984\\
		\bottomrule[1.5pt]
		
	\end{tabular}
	
	\label{table:fewshot}
\end{table}
The results are shown in Table \ref{table:fewshot}. The number $0$ indicates the original model without fine tuning. Specifically, when only $2$ function pairs were utilized, the achieved AUC of $0.719$ surpassed the original model's AUC of $0.679$. Moreover, our approach exhibited exceptional performance with just $512$ samples, reaching comparable results to models trained on the complete dataset. Additionally, as the training samples increased to $2048$, SimCLF demonstrated further convergence towards achieving similar performance to models trained on the entire dataset. These findings underscore the effectiveness of SimCLF in real-world scenarios where data scarcity is commonplace. 
The results are also compared with TREX model finetuned with InferSent model \citep{pei_TREX_nodate}. By using TREX finetuned with InferSent model, it achieve a maximum AUC of $0.942$ with $10,240$ labelled data.
\subsection{RQ5: Uniformity and Alignment}
The metrics, uniformity and alignment, that impact model performance is evaluated in this section. All the test are evaluated with the jTrans-SimCLF model. The models are trained using the parameters described in Section \ref{sec:setups}, and then tested on the BinaryCorp-3M dataset.
\begin{table}[h]
	\caption{Alignment and Uniformity of jTrans Fine-tined with Diffrent Model}
	\begin{tabular}{c|cc}
		\toprule 

 Model&Uniformity&Alignment\\
		\midrule 
 jTrans-zero& -0.598& 1.393\\     
 jTrans& -1.046&1.625 \\          
 jTrans-SimCLF& -0.813&0.914\\    

		\bottomrule[1.5pt]
	\end{tabular}
 \label{uniformity}
\end{table}
Table \ref{uniformity} shows uniformity and alignment of different function embedding models along with their averaged results. jTrans-zero is the model without finetuned. jTrans is the model finetuned with triplet loss. In general, models which have both better alignment and uniformity achieve better performance. In general, jTrans-SimCLF demonstrated superior alignment and uniformity. Additionally, it was observed that (1) although the pre-trained jTrans-zero model displayed good alignment, its uniformity was poor; and (2) while jTrans exhibited the best uniformity, its alignment was poor. Therefore, it can be concluded that jTrans-SimCLE has the ability to simultaneously optimize both alignment and uniformity metrics.

\subsection{RQ6: Temperature and Epoch}
The influence of some hyperparameters such as the number of training epochs and temperature is evaluated in this section.
Since both SimCLF and TREX use the same pre-trained model, the difference at the fine-tuning phase is compared. The fine-tuning data for the TREX model contains approximately $89,809$ function pairs. SimCLF model requires only $22,453$ functions. Although SimCLF model requires fewer samples, it converges significantly faster and better accuracy can be obtained after only about 2 rounds. As shown in the Table \ref{converge}, the TREX model converges much slower. After 20 training rounds, the average AUC is still lower than that obtained by training SimCLF for 2 rounds. The results show that by avoiding the over-training problem caused by mislabelling, SimCLF converges significantly faster with less data. 
\begin{table}[h]
	\caption{The Performance of Different Training Epochs}
	\begin{tabular}{ccccc}
		\toprule 
		Number of epochs&2&10&20&40\\
		\midrule 
		AUC score of TREX& 0.679&0.719&0.737&0.984\\
		AUC score of SimCLF& 0.917&0.989&0.990&0.992\\
		\bottomrule[1.5pt]
		\label{converge}
	\end{tabular}
\end{table}

\begin{figure}[h]
	\centering 
	\includegraphics[scale=0.4]{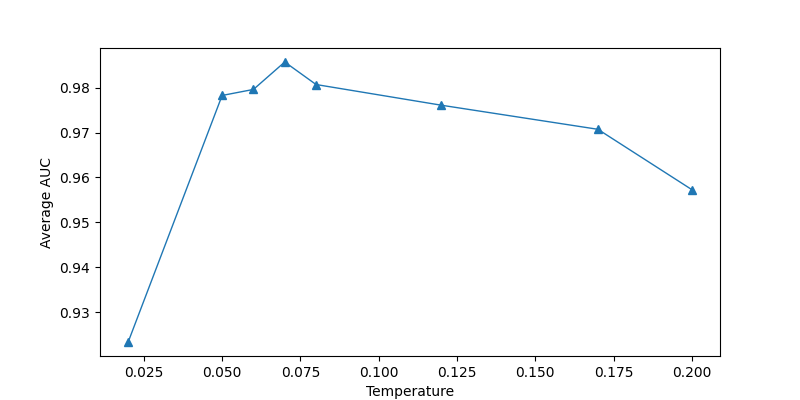} 
	\caption{The influence of different temperatures on the loss function.} 
	\label{figtemp}
\end{figure}

The importance of the temperature $\tau $ is also evaluated. The temperature $\tau $ in the loss (Equation (\ref{eq_tau}) ) is used to control the smoothness of the distribution. The smaller the $\tau $ is the sharper the distribution will be. Figure \ref{figtemp} shows the AUC at different temperatures. The best performance is obtained when the temperature is set to $0.07$.
\section{Discussion and Future Works}
Inspired by the advances in the field of NLP, many people \citep{pei_TREX_nodate}\citep{massarelli_safe_2019} are using NLP-related techniques to solve problems in the field of assembly language. Although many people use the BERT model to generate embeddings \citep{pei_TREX_nodate}\citep{li_palmtree_2021}, they all ignore the robustness of learned embeddings. In this paper, unsupervised contrastive learning is used to transform the function-level embedding learning into an instance discrimination problem. The difficulty of accurately labelling function pairs is avoid, and SimCLF achieves SOTA achievements on multiple datasets. At the same time, SimCLF shows some characteristics different from images, such as very high accuracy without project heads which merits further investigation. Other advances in contrastive learning \citep{he_momentum_nodate} are also worth being applied in BCSD fields. 

\section{Conclusion}

In this paper, we introduce SimCLF, a self-supervised contrastive learning framework designed to generate embeddings for binary functions, with the aim of facilitating downstream tasks. Notably, SimCLF stands out due to its simplicity, as it does not necessitate additional intricate structures and can be seamlessly integrated with any encoder. Moreover, we present the pioneering application of unsupervised contrastive learning techniques to pre-trained models within the BCSD domains.

To substantiate the efficacy of SimCLF, extensive experiments were conducted across diverse datasets, revealing compelling results. SimCLF consistently achieved state-of-the-art performance. Furthermore, our framework exhibited remarkable resilience in the few-shot setting, demonstrating its robustness even when faced with limited training data. As proponents of open research and knowledge sharing, we have made SimCLF codes publicly available, with the intention of inspiring among fellow researchers. We hope that our contributions will motivate further exploration and innovation in related domains.


\bibliographystyle{cas-model2-names}


\bibliography{mybibfile}


\end{document}